# Sulfur in the Giant Planets, their Moons, and Extrasolar Gas Giant Planets

Katharina Lodders and Bruce Fegley

Department of Earth, Environmental, & Planetary Sciences and The McDonnell Center for Space Sciences, Washington University, St Louis MO 63130 USA, lodders@wustl.edu, bfegley@wustl.edu

## Abstract

We review the sulfur chemistry of the gas giant planets and their moons where sulfur compounds have been observed. Hydrogen sulfide $H_2S$ is the major S-bearing gas in the upper atmospheres of the giant planets and is removed from their observable atmospheres by condensation into cloud layers ($NH_4SH$ on all four planets and additionally $H_2S$ ice on Uranus and Neptune). Any remaining $H_2S$ at higher altitudes is destroyed photochemically.

Among the moons – or any other object in the solar system - Io is the world dominated by sulfur. We summarize the sulfur cycle on Io and how Io's pyrovolcanism is spreading sulfur across the Jovian system. Implantation of sulfur into the icy surfaces of the other Galilean moons via magnetospheric transfer and radiolysis are the dominant processes affecting the sulfur chemistry on their icy surfaces. On the icy worlds, we are literally looking at the top of the icebergs. Subsurface liquid salty bodies reveal themselves through cryovolcanism on Europa, Ganymede, and Enceladus, where salt deposits are indicated. Subsurface oceans are suspected on several other moons. We summarized the sulfur cycle for the icy Galilean moons. The occurrence of sulfates can be explained by salt exchange reactions of radiolytically produces $H_2SO_4$ with brine salts such a carbonates and halides, or from a subsurface ocean that has become acidified by uptake of $H_2SO_4$ leaked from ice over time. In the primordial oceans of the moons that accreted with high ice rock rations, sulfur is expected the form of sulfide and bisulfide anions together with $H_2S$ in aqueous solution Considering various cosmochemical constraints, we suggest that in addition to pyrrhotite, tochilinite and green rusts could be important sulfide bearing compounds that reside together with hydrous silicates such as serpentine, and magnetite on the sea floors. In nitrogen-carbon rich worlds such as Titan, sulfides such as $NH_4SH$ and possibly thiazyl compounds could be important, and sulfates are unstable. Nothing is known about the sulfur chemistry on the Uranian and Neptunian moons.

## Introduction

Sulfur is the 10[th] most abundant element in the Sun, the solar system, and in the Milky Way Galaxy (Palme et al. 2014) and is the second element in the chalcogen group (O, S, Se, Te, Po, Lv) of the periodic table. Sulfur displays a wide range of oxidation states ranging from -2 to +6 in its chemistry and is found as gaseous, icy, and rocky phases in different solar system materials. In this chapter we focus on sulfur chemistry in the gas giant planets of our solar system, their major satellites, and extrasolar gaseous planets orbiting other stars. Although we focus on observational data, we also include theoretical calculations where appropriate and where observations are lacking.

## Background information about the giant planets

Jupiter, Saturn, Uranus, and Neptune are the four giant planets in our solar system. The masses and radius of the four giant planets are several times that of the Earth ($M_E = 5.97 \times 10^{24}$ kg and $R_E = 6,371$ km). The large mass, large size, and low bulk density of Jupiter (318 $M_E$, 11 $R_E$, 1.33 g cm$^{-3}$) and Saturn (95.2 $M_E$, 9.5 $R_E$, 0.687 g cm$^{-3}$) show they are primarily composed of hydrogen and helium with Saturn being more enriched than Jupiter in elements heavier than helium (self-compression effects are smaller for Saturn than Jupiter, hence a larger heavy element fraction is required to give the observed density). The smaller size, smaller mass, and higher bulk density of Uranus (14.5 $M_E$, 4.0 $R_E$, 1.32 g cm$^{-3}$) and Neptune (17.2 $M_E$, 3.9 $R_E$, 1.64 g cm$^{-3}$) – relative to those of Jupiter and Saturn – show they are primarily composed of elements heavier than He (e.g., O, C, Ne, N, Mg, Si, Fe, S). The observations and interior structure models lead to Jupiter and Saturn being called gas giant planets with Uranus and Neptune known as ice giant planets.

During formation of the giant planets the two most abundant elements (H and He) plus Ne accreted as gas, the next most abundant elements O, C, N, and Ne condensed as ices (e.g., $H_2O$, CO, $CO_2$, $CH_4$, $N_2$, $NH_3$, Ne, and/or clathrate hydrates), while the less abundant elements Mg, Si, Fe, and S accreted as condensed rock containing $MgO + SiO_2 + Fe + FeS$, and variable amounts of oxidized iron. The rocky component is probably similar to rocky material in carbonaceous chondrites. Organic matter like that found in carbonaceous chondrites may also have been present. This division into gas, ice, rock, and organics is somewhat arbitrary and one may argue that sulfur should be partitioned between rock, organics, and ice because FeS is ubiquitous in meteorites, sulfur is present in organic matter in carbonaceous chondrites (Hayes 1967), and many sulfur gases are observed degassing from comets (e.g., $H_2S$, $H_2CS$, OCS, $SO_2$, $CH_3SH$, $C_2H_6S$) (Crovisier et al 1991, Calmonte et al. 2016). Teanby et al. (2020) present an alternative view of Uranus and Neptune as rocky giant planets. Because of the larger enrichment of heavy elements on Uranus and Neptune relative to Jupiter and Saturn, perhaps ice + organic + rock (IOR) giant planets is a better descriptor of Uranus and Neptune.

There are no observable surfaces in any of the giant planets – their radii refer to the one bar atmospheric level – and as far as it is known their atmospheres extend to many thousands of kilometers below the visible cloud tops. Contemporary ideas postulate metallic H + He layers inside Jupiter (at 0.8× its radius $R_J$) and Saturn (at 0.55× $R_S$) and ionic layers ($H_3O^+$, $NH_4^+$, $OH^-$) inside Uranus (at 0.7× $R_U$) and Neptune (at 0.8× $R_N$). Rocky cores may exist at the centers of the planets or there may be a gradual increase of mean atomic mass with depth inside the metallic and ionic layers (e.g., see Guillot 2005, Bailey and Stevenson 2021).

Earth – based and spacecraft infrared (IR) observations show that Jupiter (1.7×), Saturn (1.8×), and Neptune (2.6×) emit more heat than they receive from the Sun. Unlike the other giant planets Uranus apparently has a weak internal heat source and emits at most only 14% more heat than it receives from the Sun. The internal heat flux emitted by the giant planets is transported via convection, which means that their lower atmospheres are adiabatic with temperature gradients given by

$$\frac{dT}{dz} = -\frac{\mu g}{C_P} = -\frac{g}{c_p}$$

$$\ln\left(\frac{P}{P_0}\right) = \frac{C_P}{R}\ln\left(\frac{T}{T_0}\right)$$

The symbols in the equations stand for the following: altitude is z (km), $\mu$ the mean molar mass of atmospheric gas, $C_P$ the average constant pressure heat capacity (J $mol^{-1}$ $K^{-1}$) and $c_p$ the average specific heat (J $g^{-1}$ $K^{-1}$) of atmospheric gas, and g the average gravitational acceleration. The dT/dz values are about $1 - 2.3$ K $km^{-1}$ temperature increase with increasing depth below the one bar level in the atmosphere and decrease with increasing depth because heat capacity (specific heat) increases with temperature. For reference at 1 bar in Jupiter's atmosphere, $z \equiv 0$, T = 165 K, dT/dz = 2.3 K $km^{-1}$, g = 25.4 m $s^{-2}$, $\mu$ = 2.3 g $mol^{-1}$, $C_P$ = 25.5 J $mol^{-1}$ $K^{-1}$, and $c_p$ = 11.1 J $g^{-1}$ $K^{-1}$ (Lodders and Fegley 2011).

We briefly note that condensation cloud formation releases heat back into the surrounding atmosphere and gives a lower ("wet") temperature gradient in the cloud-forming region, which reverts back to the cloud-free value at the cloud tops. This situation occurs with each condensation cloud layer be it liquid iron droplets, silicate rock, or $NH_4SH$ particles. Weidenschilling and Lewis (1973) describe derivation of the "wet" adiabatic temperature gradient and Li et al. (2018) generalize the derivation for a multicomponent condensate system with large amounts of condensates.

Voyager spacecraft measurements show the upper atmospheres of all four planets have temperature profiles that switch from radiative in the stratospheres to convective in the troposphere with increasing depth. The radiative – convective boundary occurs at the tropopause which is at 140 millibar (mbar), 104 K on Jupiter; 80 mbar, 85 K on Saturn; 100 mbar, 54 K on Uranus; and 200 mbar, 52 K on Neptune (Visscher 2022). The Galileo probe into Jupiter's atmosphere measured an adiabatic temperature profile down to its maximum operating depth of about 22 bar, 428 K (Seiff et al 1998). For illustration, extrapolation of the Jovian adiabatic P – T profile to higher temperature (and greater depth) gives pressures of about 400 bar at 1000 K and about 5.5 kilobar at 2000 K. The troposphere of Saturn, Uranus, and Neptune are also adiabatic to great depths and follow P – T profiles that are shifted to higher pressure than the Jovian adiabat at the same temperature. The high temperatures and pressures favor chemical equilibrium in the deep troposphere of the four planets.

Table 1 shows the major gases observed in the atmospheres of the giant planets correspond to those expected in a solar (or near solar $H_2$ – rich) composition atmosphere at chemical equilibrium. Lewis (1969b) first made this point, which is fundamental for understanding atmospheric chemistry of the gas and ice giants in our solar system and in exoplanetary systems. The dominance of $H_2$ and high temperatures and pressures in the deep troposphere of the gas giants favors hydrides such as $H_2O$, $CH_4$, $NH_3$, and $H_2S$ over their oxidized counterparts (e.g., $O_2$, $O_3$, CO, $CO_2$, $N_2$, $NO_x$, $SO_2$) found in the atmospheres of Venus, Earth, and Mars, e.g., via net thermochemical reactions such as

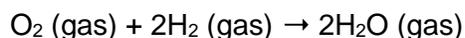

$O_2$ (gas) + $2H_2$ (gas) → $2H_2O$ (gas)

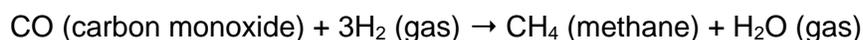

CO (carbon monoxide) + $3H_2$ (gas) → $CH_4$ (methane) + $H_2O$ (gas)

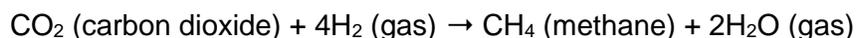

$CO_2$ (carbon dioxide) + $4H_2$ (gas) → $CH_4$ (methane) + $2H_2O$ (gas)

N$_2$ (gas) + 3H$_2$ (gas) → 2NH$_3$ (ammonia)

SO$_2$ (sulfur dioxide) + 3H$_2$ (gas) → H$_2$S (hydrogen sulfide) + 2H$_2$O (gas)

Table 1 includes Mg, Si, and Fe, which are the 7$^{th}$ – 9$^{th}$ most abundant elements in solar composition material (Lodders 2020). These elements are not observed in the atmospheres of the giant planets because they are sequestered in rock and metal deep inside the giant planets, as predicted by chemical equilibrium (e.g., see Lewis 1969a, Barshay and Lewis 1978, Fegley and Lodders 1994):

2H$_2$O (gas) + SiH$_4$ (silane) → 4H$_2$ (gas) + SiO$_2$ (in rock)

Mg (gas) + H$_2$O (gas) → H$_2$ (gas) + MgO (in rock)

Fe (gas) → Fe (metal)

However, Visscher et al. (2010) show Si-, Mg-, and Fe-bearing gases may be observable in gas giant exoplanets and brown dwarfs that are hotter than is Jupiter's atmosphere at the same total pressure.

**Table 1.** Solar elemental abundances compared to giant planet atmospheres

| Element | Atoms[a] | Species | "Solar"[b] | Jupiter | Saturn | Uranus | Neptune |
|---|---|---|---|---|---|---|---|
| H | 2.52×10$^{10}$ | H$_2$ | 83.4% | 86.4% | 88% | 82.5% | 80% |
| He | 2.51×10$^9$ | He | 16.6% | 13.6% | 12% | 15.2% | 14.9%[c] |
| O | 2.51×10$^7$ | H$_2$O | 898 | 567 | 2-20 ppb | – | – |
| C | 9.12×10$^6$ | CH$_4$ | 604 | 2743 | 4700 | 2.3% | 2 – 5%[c] |
| Ne | 4.37×10$^6$ | Ne | 289 | 21 | – | – | – |
| N | 2.19×10$^6$ | NH$_3$ | 145 | 340[d] | 160 | 170[f] | 470[g] |
| Mg | 1.03×10$^6$ | rock | 68 | – | – | – | – |
| Si | ≡10$^6$ | rock | 66 | – | – | – | – |
| Fe | 8.72×10$^5$ | metal | 58 | – | – | – | – |
| S | 4.37×10$^5$ | H$_2$S | 29 | 103 | <0.4[e] | 870[f] | 1300[g] |

[a]Protosolar elemental abundances (Lodders 2020).
[b]Hypothetical solar composition giant planet atmosphere. The H$_2$O value is reduced from the oxygen solar abundance by about 20% due to the amount of oxygen in rock (MgO + SiO$_2$) and is not simply total oxygen as water. The values for gases are percent by volume for H$_2$ and He, parts per million by volume (ppmv) for all other gases except H$_2$O on Saturn, which is parts per billion by volume (ppbv).
[c]Moses et al. 2016, [d]Moeckel et al. 2023, [e]Upper limit from Owen et al 1977, [f]Molter et al 2021, [g]Tollefson et al 2021

The abundances and vertical profiles of the major hydride gases on Jupiter, Saturn, Uranus, and Neptune are controlled by three factors: (1) Enrichment of heavy elements relative to their solar abundances, (2) Photochemical destruction of hydrides in giant planet stratospheres where short wavelength UV light penetrates, and (3) Condensation cloud formation in the convective tropospheres. We focus on Jupiter, where the most data are available, to illustrate the importance of these three factors and mention the other giant planets as appropriate.

Methane is observable by Earth – based spacecraft IR spectroscopy on all four giant planets. The $CH_4$ mole fractions in the four giant planets illustrates heavy element enrichment over solar composition. The $CH_4$ mole fraction is 2743 parts per million by volume (ppmv) and it is the dominant carbon-bearing gas on Jupiter with its photoproduct $C_2H_6$ being a distant second at 5.8 ppmv. Table 1 shows the $CH_4$ mole fraction on Jupiter is about 4.5× higher than in a solar composition atmosphere (i.e., $2743/604 \simeq 4.5$). Saturn, Uranus, and Neptune have even higher $CH_4$ mole fractions with enrichments of 7.8×, 38×, and 33 – 83× solar, respectively.

Photochemical destruction of hydrides is exemplified by UV sunlight driven $CH_4$ photolysis high in the Jovian stratosphere that converts it into other hydrocarbons (mainly $C_2H_6$ and $C_2H_2$) via the net photochemical reactions:

$2CH_4$ (methane) $\rightarrow C_2H_6$ (ethane) $+ H_2$ (gas) $\hspace{4cm} \lambda < 145$ nm

$2CH_4$ (methane) $\rightarrow C_2H_2$ (acetylene) $+ H_2$ (gas) $+ 4H$ (gas) $\hspace{2cm} \lambda < 145$ nm

Ethane, $C_2H_2$, and other hydrocarbons produced by $CH_4$ photolysis are recycled back to methane by thermochemical reactions in the deep, unobservable atmospheres of Jupiter (and the other three giant planets), e.g., via the reactions:

$H_2$ (gas) $+ C_2H_6$ (ethane) $\rightarrow 2CH_4$ (methane)

$3H_2$ (gas) $+ C_2H_2$ (acetylene) $\rightarrow 2CH_4$ (methane)

Ammonia is photolyzed by UV photons at 160 – 235 nanometers in the vicinity of the Jovian tropopause via the net photochemical reactions:

$2NH_3$ (ammonia) $\rightarrow N_2 + 3H_2$ (gas) $\hspace{5cm} \lambda = 160 – 235$ nm

$2NH_3$ (ammonia) $\rightarrow N_2H_4$ (hydrazine) $+ H_2$ (gas) $\hspace{3cm} \lambda = 160 – 235$ nm

Methane, $H_2$, and He shield $NH_3$ from photons shortward of 160 nm and $NH_3$ does not absorb photons longward of 235 nm. Formation of $N_2$ dominates and hydrazine, which condenses into aerosols, is a minor product. Downward transport and reduction by $H_2$ in the deep tropospheres of Jupiter and Saturn reforms $NH_3$ from its photoproducts. Ammonia photolysis is unimportant on Uranus and Neptune because their upper atmospheres are so cold that $NH_3$ is condensed into $NH_3$ (ice) clouds far below their stratospheres where UV photons at <230 nm penetrate (Lodders and Fegley 2011).

Photochemistry of water vapor is unimportant on the giant planets in our solar system because $H_2O$ is condensed as liquid water clouds at greater depth than the $NH_3$ ice clouds and no UV photons (with $\lambda < 212$ nm) that can photolyze $H_2O$ are present this deep in giant planet atmospheres. Earth-based IR spectroscopy reveals only trace amounts of water vapor in their observable atmospheres. The Galileo entry probe observed about 567 ppmv water vapor below the water clouds on Jupiter (Niemann et al 1998, Wong et al 2004), which is 0.6× the amount expected in a solar composition atmosphere (Table 1). Microwave sounding by the JUNO orbiter indicates about 2500 ppmv of water in the 0.7 – 30 bar region of Jupiter's atmosphere (Li et al. 2020), which is about 2.8× the solar value of 898 ppmv $H_2O$ in Table 1.

Hydrogen sulfide was discovered on Jupiter by the mass spectrometer on the Galileo entry probe (Niemann et al 1998) and is present at a mole fraction of about 103 ppmv (Wong et al. 2004). This is about 3.6× the solar value of 29 ppmv sulfur (Lodders 2020).

An earlier report of $H_2S$ dredged up by impacts of comet Shoemaker – Levy 9 (SL9) during July 1994 is moot as described by Lellouch (1996). However, in order of decreasing abundance, the sulfur-bearing gases OCS, $S_2$, CS, and $CS_2$ were produced by the SL9 impacts (Lellouch 1996). Rapid quenching of high temperature driven shock chemistry is the probable origin of these species, which form either in regions containing more (OCS) or less ($S_2$, CS, $CS_2$) water vapor (Zahnle et al. 1995). No $SO_2$ or SO were observed from the SL9 impacts. Their absence limits the amount of water in the shock heated cometary + Jovian gas mixture because $SO_2$ is the dominant S – bearing gas produced from a "wet" comet with solar ratios of $C:N:S:H_2O$ (Tables 1 & 2 of Zahnle et al. 1995). We return to the topic of $SO_2$ later.

Hydrogen sulfide, like $H_2O$ and $NH_3$, is removed from Jupiter's atmosphere by condensation cloud formation, but not as pure $H_2S$ ice. Instead, it reacts with $NH_3$ to form $NH_4SH$, ammonium hydrosulfide, which we discuss in more detail in the next section. But first we complete our discussion of photochemistry.

On Jupiter, $H_2S$ photolysis occurs in the vicinity of the $NH_4SH$ cloud and is driven by solar UV photons in the 235 – 270 nm wavelength range (Lewis and Prinn 1984). The initial step is production of H atoms and SH (mercapto) radicals,

$$H_2S + h\nu \rightarrow SH + H \qquad \lambda = 235 - 270 \text{ nm}$$

Mercapto radicals are also produced via

$$H + H_2S \rightarrow H_2 + SH$$

The two major reactions reforming $H_2S$ are

$$H + SH + M \rightarrow H_2S + M$$

$$SH + SH \rightarrow H_2S + S$$

The M stands for a collision partner, which is statistically one of the most abundant gases in the atmosphere, e.g., $H_2$, He, $CH_4$, and so on. The collision partner is needed to dissipate the excess energy released by H – SH bond formation.

Recombination of H atoms to form $H_2$ is a slow process because three bodies are involved

$$H + H + M \rightarrow H_2 + M$$

The principal loss processes for mercapto radicals are reformation of $H_2S$, production of monatomic S, formation of $S_2$ vapor, and the production of $H_2S_2$ (hydrogen disulfide),

$$H + SH + M \rightarrow H_2S + M$$

$$SH + SH \rightarrow H_2 + S_2$$

$$SH + SH \rightarrow H_2S + S$$

$$SH + SH + M \rightarrow H_2S_2 + M$$

Subsequent reactions lead to even numbered sulfur allotropes up to $S_8$, which condenses, e.g.,

$$SH + S \rightarrow H + S_2$$

$$S_2 + S_2 + M \rightarrow S_4 + M$$

$$S_2 + S_4 + M \rightarrow S_6 + M$$

$$S_2 + S_6 + M \rightarrow S_8 + M$$

$$S_8 = S_8 \text{ (solid)}$$

Odd sulfur allotropes are produced by elementary reactions such as

$$S + S_2 + M \rightarrow S_3 + M$$

$$S + S_4 + M \rightarrow S_5 + M$$

$$S + S_6 + M \rightarrow S_7 + M$$

Loss processes include solar UV photolysis of sulfur allotropes and disproportionation reactions, e.g., three possible reactions are

$$S_3 + h\nu \rightarrow S_2 + S$$

$$S_4 + h\nu \rightarrow S_2 + S_2$$

$$S_3 + S_7 \rightarrow S_4 + S_6$$

At present there are no observations of HS, $H_2S_2$, or the sulfur allotropes on any of the giant planets and the $H_2S$ photochemical scheme remains theoretical. The measurement of these reactive species at low concentrations is difficult to say the least. However, $S_2$ was measured by the scanning spectrophotometer on the Venera 11/12 entry probes (Moroz et al. 1980) and the abundances of other sulfur allotropes were calculated assuming chemical equilibrium by San'ko (1980). It may be possible to use a magnetic sector mass spectrometer on an entry probe, like that on the Pioneer Venus sounder probe (Hoffman et al. 1980, Ren et al. 2018), to measure the more abundant species produced by $H_2S$ photolysis and distinguish them from potential isobaric interferences.

 Solar UV photolysis of the $NH_4SH$ cloud particles is also possible. Lebofsky and Fegley (1976) showed UV (220 – 300 nm) irradiated $NH_4SH$ frost layers produced visible chromophores absorbing around 600 nm within a few hours. They suggested the chromophores may be $S_5 - S_7$ diradicals. Loeffler and colleagues studied reflection spectra of proton irradiated $NH_4SH$ and observed similar spectral features (Loeffler et al. 2015, 2016, Loeffler and Hudson 2018). However experimental studies done at 200 – 260 K, spanning those expected at the $NH_4SH$ cloud base in the atmospheres of Jupiter and Saturn show little absorption at 600 nm and appear colorless (e.g., see Figure 1 of Loeffler et al. 2016).

Photochemical production of S – bearing organic compounds, exemplified by the most abundant expected photoproduct – methyl mercaptan $CH_3SH$ – is negligible because of the different altitude regions in which $CH_4$ and $H_2S$ are photolyzed. Lewis and Fegley (1979) demonstrated this point by discussing $CH_3SH$ formation from "hot" hydrogen atoms in Jupiter's atmosphere.

The hot H atoms are produced by $H_2S$ photolysis in the 235 nm (5.3 eV) to 270 nm (4.6 eV) range and have excess energy of 0.7 – 1.4 eV (electron volts) relative to the H – SH bond strength of 3.9 eV. Methyl mercaptan is produced via the elementary reaction sequence:

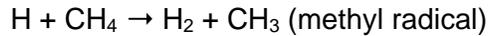

$$H + CH_4 \rightarrow H_2 + CH_3 \text{ (methyl radical)}$$

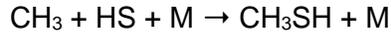

$$CH_3 + HS + M \rightarrow CH_3SH + M$$

Their estimated $CH_3SH$ mole fraction is about $10^{-16}$, about 5,000 times smaller than expected from quenching of $CH_3SH$ in upwelling gas parcels (Barshay and Lewis 1978). The main conclusion of Lewis and Fegley (1979) is that organosulfur compounds are expected to be irrelevant in the atmosphere of Jupiter and by analogy the atmospheres of the other three giant planets.

## Condensation clouds in atmospheres of the gas giant planets

### Jupiter and Saturn

Temperatures in the upper tropospheres of Jupiter and Saturn are low enough to enable condensation clouds of $H_2O$ (ice or liquid) and $NH_3$ (ice) at altitudes where the partial pressures of $H_2O$ and $NH_3$ intersect the vapor pressure curves of water ice and ammonia ice. At and above these altitudes, generally denoted as the condensation cloud bases, the atmospheric abundances of $H_2O$ and $NH_3$ decrease exponentially with decreasing temperature because they are constrained to follow the vapor pressure curves. Temperatures in the upper tropospheres of Uranus and Neptune are even lower and $CH_4$ (ice) condensation also occurs.

Cloud condensation also occurs lower in the hot, deep atmospheres of the gas giant planets as first explored by Lewis (1969b) and subsequently by Barshay and Lewis (1978), Fegley and Lodders (1994), and Lodders (1999). Condensates in a gravitationally bound planetary atmosphere that form directly from the gas at high temperatures (primary condensates) settle due to gravity into cloud layers (sometimes called "rainout"). This process removes the highest temperature condensates from the gas overlying the condensate cloud. The least abundant constituent element limits the amount of condensate. The primary condensates are not present to react with cooler, lower pressure gas at higher altitudes above the primary condensate clouds to form secondary condensates. Thus, the primary condensate clouds are out of equilibrium with the gas in the overlying atmosphere. In contrast, condensates in protoplanetary disks (e.g., solar nebula) or stellar outflows can remain dispersed in the gas and react with it to form secondary condensates during cooling. This important distinction accounts for many of the differences between chemistry of planetary atmospheres and protoplanetary accretions disks having solar, or near solar, composition (e.g., Lodders 1999, Lodders and Fegley 2002, Lodders and Fegley 2006).

Troilite, FeS, formation is a relevant example here. Troilite cannot form in planetary atmospheres in contrast to what is well known for protoplanetary disks such as the solar nebula. There troilite forms around 700 K via the pressure-independent reaction (e.g., Larimer 1967, Grossman 1972, Lewis 1972, Wai and Wasson 1977, Sears 1978, Lodders 2003)

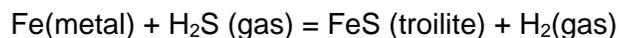

$$Fe(metal) + H_2S \text{ (gas)} = FeS \text{ (troilite)} + H_2(gas)$$

that consumes all $H_2S$ gas because the Fe/S elemental abundance ratio equals 2 in solar composition matter (Lodders 2020). However, condensate cloud models predict formation of liquid Fe clouds deep in the Jovian atmosphere at temperatures above 2500 K (e.g., Figure 1 in Visscher et al. 2010). No Fe metal or gas is left in Jupiter's higher atmosphere by 700 K where troilite can form. As a result, no FeS condenses and $H_2S$ is left in the atmosphere. The Galileo probe mass spectrometer (GPMS) detected $H_2S$ at 3.6X times the solar $S/H_2$ ratio in Jupiter's atmosphere (Niemann *et al.* 1998, Wong et al. 2004).

Hydrogen sulfide condenses out of the cool upper tropospheres of the gas giants but in a slightly different manner than $NH_3$ or $H_2O$. As suggested by Wildt (1937) and quantified by Lewis (1969a) and Weidenschilling and Lewis (1973), $H_2S$ undergoes chemical reaction with $NH_3$ and condenses as the salt ammonium hydrosulfide $NH_4SH$:

<div align="center">

$NH_3$ (gas) + $H_2S$ (gas) = $NH_4SH$ (solid)        condensation

</div>

The equilibrium constant $K_P$ in terms of the $NH_3$ and $H_2S$ partial pressures is given by

$$K_P = \frac{a_{NH_4SH}}{P_{NH_3}P_{H_2S}} = \frac{a_{NH_4SH}}{X_{NH_3}X_{H_2S}P_T^2}$$

The $a_i$ term is the thermodynamic activity of $NH_4SH$, which is unity when it becomes stable and condenses. The activity of $NH_4SH$ is less than unity above its condensation temperature and is fixed at unity once it condenses. The $X_i$ terms are the mole fractions of the two gases and the $P_T$ term is the total pressure at some level in the atmosphere of a gas giant planet.

Isambert (1881, 1882) and Walker and Lumsden (1897) measured the total vapor pressure $P_{vap}$ over $NH_4SH$ solid as a function of temperature from 4.2 – 44.4 °C. The different measurements give slightly different values for the vapor pressure. Two other groups measured vapor pressure data at a single temperature (20.0 °C – Magnusson 1907; 0.0 °C – Scheflan and McCrosky 1932).

Kelley (1937) analyzed the data of Walker and Lumsden (1897). Assuming that $NH_4SH$ dissociates to $NH_3$ and $H_2S$ without any $NH_4SH$ gas then the total vapor pressure is simply the partial pressure sum:

$$P_{vap} = P_{NH_3}^{vap} + P_{H_2S}^{vap}$$

The equilibrium constant for $NH_4SH$ vaporization ($K_{vap}$) is then:

$$K_{vap} = \frac{P_{vap}^2}{4} = P_{NH_3}^{vap} \times P_{H_2S}^{vap}$$

The product of the $NH_3$ and $H_2S$ equilibrium partial vapor pressures is the inverse of $K_P$, i.e.,

$$K_{vap} = \frac{1}{K_P} = P_{NH_3}^{vap} \times P_{H_2S}^{vap}$$

The assumption that $NH_4SH$ vaporizes to $NH_3$ and $H_2S$ is supported by the work of Scheflan and McCrosky (1932). Vaporization of other ammonia salts also suggest dissociation upon evaporation; e.g., Knudsen effusion mass spectrometric (KEMS) measurements by Goldfinger and Verhaegen (1969) suggested <0.5% $ND_4Cl$ gas in the saturated vapor over $ND_4Cl$ solid. Conversely, Hänni et al. (2019) studied sublimation of ammonium chloride $NH_4Cl$ and ammonium formate $NH_4HCO_2$ by KEMS and concluded both substances underwent dissociative vaporization. Altwegg et al. (2022) surmised $NH_4SH$ in grains of comet 67P/Churyumov-Gerasimenko from the simultaneous outbursts of $NH_3$ and $H_2S$. They did not report any evidence for $NH_4SH$ gas. We conclude that the presence of small amounts of $NH_4SH$ gas cannot be ruled out but seems unlikely.

Kelley (1937) derived the vapor pressure equation:

$$\log K_{vap} = 14.89 - \frac{4,732}{T}$$

Lewis (1969a) gave a similar equation from his fit to the $NH_4SH$ vapor pressure data in the 1961 edition of the Handbook of Chemistry and Physics:

$$\log K_{vap} = 14.82 - \frac{4,705}{T}$$

We use Kelley's (1937) vapor pressure equation in what follows.

Rearranging the equilibrium constant expression, taking logarithms, and solving for the temperature where the activity of $NH_4SH$ is unity gives

$$log\, a_{NH_4SH} = 0 = logK_P + logX_{NH_3} + logX_{H_2S} + 2logP_T$$

Substituting values from Kelley (1937) for the equilibrium constant, assuming all N and S are in $NH_3$ and $H_2S$, and taking values from Lodders (2020) for the solar abundances of N and S, we get

$$T_{cond}(NH_4SH) = \frac{-4,732}{[-14.89 - 8.378 + 2logP_T]}$$

Table 2 lists the condensation temperatures calculated with this equation. Using Lewis' vapor pressure equation gives an insignificant difference in the computed condensation temperatures.

Table 2. $NH_4SH$ condensation temperatures – solar composition material

| Log $P_T$ /bar | -4 | -3 | -2 | -1 | 0 | 1 | 2 |
|---|---|---|---|---|---|---|---|
| $T_{cond}$ (K) | 151 | 162 | 174 | 187 | 203 | 222 | 246 |

The $NH_4SH$ condensation temperatures vary by less than a factor of two over six orders of magnitude in total pressure: from 151 K at $10^{-4}$ bar to 246 K at $10^2$ bar. Linear least squares fit to the $NH_4SH$ condensation temperatures in Table 2 gives the equation

$$\frac{10,000}{T_{cond}(NH_4SH)} = 49.18 - 4.225logP_T$$

The slight difference from the equation given by Visscher et al (2006) arises from the changes in solar abundances since Lodders (2003).

The $NH_4SH$ condensation temperature is the cloud base temperature (about 220 K on Jupiter). Ammonium hydrosulfide cloud formation continues to lower temperature and higher altitude until the less abundant of the two reactants is exhausted. Hydrogen sulfide is less abundant than $NH_3$ in a solar composition atmosphere because the N/S solar abundance ratio is five (see Table 1). If nitrogen and sulfur are equally enriched relative to solar composition in a planetary atmosphere the $NH_3/H_2S$ abundance ratio remains five. In this case $H_2S$ is exhausted while $NH_3$ remains in the gas. The $H_2S$ partial pressure in equilibrium with $NH_4SH$ cloud particles and $NH_3$ gas is given by rearranging the equilibrium constant expression. Conversely if $NH_3$ were less abundant than $H_2S$ – as may be the case on Uranus and Neptune as discussed below – the $NH_3$ partial pressure in equilibrium with $NH_4SH$ cloud particles and $H_2S$ gas can also be computed from the equilibrium constant expression.

The removal of $H_2S$ by $NH_4SH$ condensation and $H_2S$ photolysis in the same altitude range are the reasons why Earth-based IR spectroscopy could not see $H_2S$ in Jupiter's atmosphere (Bézard et al. 1983, Larson et al. 1984) or Saturn's atmosphere (Owen et al. 1977).

The results in Table 2 are for a solar composition atmosphere, but they are easily modified to consider variable enrichments (depletions) of $NH_3$ and $H_2S$ relative to solar composition by inserting metallicity terms for nitrogen and sulfur:

$$\frac{10{,}000}{T_{cond}(NH_4SH)} = 49.18 - 4.225\left(logP_T + 0.5\left[\frac{N}{H}\right] + 0.5[\frac{S}{H}]\right)$$

The [N/H] and [S/H] terms are enrichments (depletions) relative to solar composition on a logarithmic scale, i.e., an enrichment of 2× solar sulfur gives [S/H] = 0.30.

Before proceeding we briefly mention some minor S-bearing gases and condensates expected in the deep atmospheres of the giant planets. Hydrogen sulfide thermally dissociates to H atoms and mercapto radicals with increasing temperature,

$$H_2S = H + SH$$

At higher temperatures the mercapto radicals can dissociate to S and H atoms. However, at a given temperature, increasing pressure forces recombination of H + S back to SH and of H + SH back to $H_2S$ and $H_2S$ remains the dominant S-bearing gas throughout Jupiter's atmosphere, at least for T ≤ 2500 K (see Figures 1 and 2 and associated text in Visscher et al 2006). As discussed by Lodders (1999) and Visscher et al. (2006), only small amounts (≈9%) of $H_2S$ are removed from the atmosphere of Jupiter by condensation of MnS (1800 K), $Na_2S$ (1370 K), and ZnS (970 K) because the solar abundance of sulfur ($4.37{\times}10^5$ atoms) is significantly greater than that of Na (57,800 atoms), Mn (9090 atoms), or Zn (1260 atoms) (Lodders 2020).

## Uranus and Neptune

Hydrogen sulfide chemistry in the observable atmospheres of Uranus and Neptune is different than on Jupiter and Saturn because of an apparent deficit of $NH_3$ in their atmospheres. Irwin et al. (2018) detected $H_2S$ (in the near IR at $1.57 - 1.59\ \mu m$) with a mole fraction of $(0.4 - 0.8)\times 10^{-6}$ at the cloud tops of Uranus. Shortly thereafter, Irwin et al. (2019) reported a probable detection (in the same wavelength region) of $H_2S$ with a mole fraction of $(1-3)\times 10^{-6}$ at the cloud tops in Neptune's atmosphere. Both groups interpreted the observed clouds as $H_2S$ ice clouds. The $H_2S$ saturation vapor pressure over $H_2S$ ice can be estimated from the vapor pressure equation of Giauque and Blue (1936),

$$logP_{H_2S}(\text{bar}) = 7.40508 - 5.1263\times 10^{-3}T - \frac{1329}{T}$$

Their equation was measured over a limited temperature range from 164.9 – 187.6 K and is increasingly in error at lower and lower temperatures. However, there are no other measurements of the vapor pressure of $H_2S$ ice.

The $H_2S$ observations imply that $H_2S/NH_3 > 1$ in the atmospheres of Uranus and Neptune, instead of being equal to the solar S/N ratio of 0.2 (Lodders 2020). Otherwise, no $H_2S$ would remain in the atmosphere above the $NH_4SH$ clouds, as is the case on Jupiter and Saturn. Microwave observations of the deep atmospheres of Uranus and Neptune – at levels far below the observable clouds – suggest $NH_3$ is significantly depleted below the solar N/H ratio (Gulkis et al. 1978, de Pater et al. 2023), while $H_2S$ is significantly enhanced above the solar S/H ratio (Molter et al. 2021, Tollefson et al. 2021).

Several mechanisms may explain the apparent nitrogen depletion. (1) Uranus and Neptune are depleted in total nitrogen to the extent that the bulk S/N ratio is greater than unity. This mechanism is plausible if the heavy element enrichments of Uranus and Neptune are due to accretion of large amounts of rocky material similar to chondritic meteorites. For example, the CI carbonaceous chondrites, which are the most nitrogen rich meteorites, contain 2500 $\mu g/g$ nitrogen and 53,600 $\mu g/g$ sulfur (Lodders 2020). The corresponding S/N molar ratio is about 9 versus the solar ratio of 0.2.

The second possible mechanism is that the observable atmospheres of Uranus and Neptune are depleted in $NH_3$ because most of their nitrogen is present as $N_2$ gas. However, the real gas chemical equilibrium calculations of Fegley and Prinn (1986) for Uranus find $NH_3$ remains dominant and suggest this explanation is unlikely.

Finally, the third possible mechanism is sequestration of $NH_3$ into lower lying water clouds or an ionic ocean (e.g., Atreya and Romani 1985, Carlson et al. 1987, Fegley and Prinn 1985, 1986, Stevenson 1984). Ammonia is highly soluble in water at room temperature. The different authors who explored this mechanism used different thermodynamic data for the pressure and temperature-dependent solubility and found different amounts of $NH_3$ dissolved in the hot, deep, water clouds in Uranus' atmosphere. It remains to be seen whether this explanation or a planetary depletion of nitrogen or some combination of both accounts for the $H_2S$ and $NH_3$ observations on Uranus and Neptune.

## Sulfur chemistry in Extrasolar Gas Giant Planets (EGPs)

Early release science with the James Webb Space Telescope (JWST) shows $SO_2$ in the atmospheres of two extrasolar gas giant planets – WASP-39b and WASP-107b (Alderson et al. 2023, Dyrek et al. 2023, Rustamkulov et al. 2023). WASP-39b is a hot-Saturn exoplanet orbiting its host G-type star at 0.0486 AU (vs. 9.5 AU for Saturn) and WASP-107b is a hot-Neptune orbiting its host K-type star at 0.055 AU (vs 30.1 AU for Neptune). The stellar irradiation on the two WASP planets is tens of thousands times higher than that for Saturn or Neptune even when the reduced flux from a K-type star is taken into account.

As discussed earlier, $H_2S$ is the major sulfur-bearing gas in the atmospheres of the giant planets in our solar system. Why is $SO_2$ observed on the WASP exoplanets, which are also $H_2$-rich objects? Tsai et al. (2023) present a photochemical scheme for production of $SO_2$ in the atmosphere of WASP-39b that answers this question:

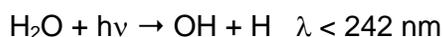
$$H_2O + h\nu \rightarrow OH + H \quad \lambda < 242 \text{ nm}$$

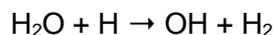
$$H_2O + H \rightarrow OH + H_2$$

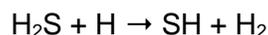
$$H_2S + H \rightarrow SH + H_2$$

$$SH + H \rightarrow S + H_2$$

$$S + OH \rightarrow SO + H$$

$$SO + OH \rightarrow SO_2 + H$$

Net photochemical reaction: $H_2S + 2H_2O \rightarrow SO_2 + 3H_2$

The presence of $SO_2$ on WASP-39b (and WASP-107b) is due to photolysis of water vapor. The primary difference between the WASP exoplanets and the giant planets in our solar system is that water condensed out of the atmospheres of Jupiter, Saturn, Uranus, and Neptune far below the atmospheric levels where $H_2O$ (or $H_2S$) absorbs photons. Water (or $H_2O$ ice) clouds do not condense on the WASP exoplanets because the observable atmospheres of WASP-39b and WASP-107b are significantly hotter than the observable atmospheres of Jupiter, Saturn, Uranus, or Neptune.

The observed $SO_2$ is the balance of photochemical production (see above) and loss via photolysis that forms sulfur monoxide SO:

$$SO_2 + h\nu \rightarrow SO + O \quad \lambda < 219 \text{ nm}$$

Possible subsequent reactions can either reform $SO_2$, or lead to condensation of elemental sulfur via reactions shown earlier when discussing $H_2S$ photochemistry on Jupiter,

$$SO + OH \rightarrow SO_2 + H$$

$$SO + h\nu \rightarrow S + O \qquad \lambda < 233 \text{ nm}$$

$$O + H_2O \rightarrow OH + OH$$

$$S + H_2S \rightarrow S_2 + H_2$$

Photochemically produced $SO_2$ and SO will be converted back to $H_2S$ at higher pressures and temperatures in the deeper atmospheres of EGPs via the net thermochemical reactions,

$$SO_2 + 3H_2 = H_2S + 2H_2O$$

$$SO + 2H_2 = H_2S + H_2O$$

Higher pressure drives reduction of $SO_2$ (and SO) to $H_2S$ and the $H_2S/SO_2$ and $H_2S/SO$ molar ratios are proportional to total pressure.

We can speculate that similar sulfur chemistry will be observed in the atmospheres of other extrasolar gas giant planets that closely orbit their parent stars unless the stellar irradiation is similar to that at the giant planets in our solar system. The key point is the presence of water vapor at atmospheric levels where stellar UV irradiation produces OH radicals. The production of $SO_2$ on EGPs is analogous to $SO_2$ production from shock heated cometary – Jovian gas mixtures (Zahnle et al. 1995). In both cases abundant water is present to provide OH radicals for oxidation of sulfur species to $SO_2$ but the energy sources are different – stellar UV photons on the EGPs and gravitational potential energy in cometary shock heated Jovian gas parcels.

# Sulfur on the moons of the giant planets

The number of currently known natural satellites of the giant planets is impressive: Jupiter 95, Saturn 146, Uranus 27, and Neptune 14. We restrict our discussion to the larger moons and those for which observations of S-bearing compounds are available.

Most of the moons considered here are differentiated rocky and icy worlds with varying degrees of geologic surface activity. One exceptional case is Io, which is the most volcanically active object in our solar system as it is strongly tidally heated by Jupiter and neighboring moons (mainly Europa). Io's surfaces are actively resurfaced by hot (up to about 1450K) lava flows rich in sulfur compounds, while at the same time having terrains with seasonal variations of $SO_2$ ice-coverage. Other moons with endogenic activity are Europa, Ganymede, and Enceladus displaying geysers and water ice cryovolcanism. Another example is Triton where seasonal cryovolcanism produces $N_2$-rich surface frost.

Most moons are essentially free of any significant denser atmospheres, here the major exception is Titan with its 1.5 bar atmosphere. The major factors influencing atmospheric generation and retention on satellites include (1) abundances of volatile elements such as H, C, N, O, Ne, Ar, and S; (2) surface temperature, (3) surface gravity, (4) volatile sources, and (5) volatile sinks. Sources include pyrovolcanism (hot molten rock), cryovolcanism (low temperature fluids such as $N_2$), and sublimation of surface ices, e.g., on Titan and Triton. Sinks include precipitation of ice(s), solar UV photolysis and atmospheric escape.

Transient and localized atmospheres may arise from impacts of micrometeoroids and larger objects on icy satellite surfaces. Also, localized atmospheres may form at pyrovolcanic plumes (Io) and cryovolcanism/geysers (Europa, Ganymede, Enceladus, Triton). Gases not lost to space may condense as frost on the surfaces if their vapor pressures at prevailing surface temperatures are low, and these cannot contribute to a global build-up of an atmosphere. Here these gases are mainly $SO_2$ and $S_2$ (Io), and $H_2O$, and $CO_2$ for other moons. At a given temperature gases with higher vapor pressures over physical mixtures of pure ices, ice solid solutions, ice-hydrates (e.g., $NH_4OH$), clathrate hydrates, or even hydrocarbon liquids are favored as atmospheric gases, which is important for $N_2$, CO, $CO_2$, $CH_4$, $NH_3$, $H_2S$, $SO_2$. Among all these potential ices, water is the most refractory one, if present, and frozen-out on the moons' surfaces.

For pyrovolcanism to operate efficiently as and serve as a source for atmospheric gases, sufficient solubilities of e.g., $H_2O$, $CO_2$, $NH_3$, $H_2S$, $SO_2$, and/or their soluble species (e.g., $OH^-$, $NH_4^+$, $N^{3-}$, $CO_3^{2-}$, $S^{2-}$, $SO_4^{2-}$) in magmas is relevant. The solubility of sulfur is long known to depend on the oxidation state of the magmas with a U-shaped variation at intermediate $fO_2$ and higher solubility as sulfate $(SO_4)^{2-}$ under oxidizing conditions and as sulfide, $S^{2-}$ under more reducing conditions, which is important for Io (Zolotov and Fegley 1999).

During cryovolcanism, atmospheric gases can come from putative (ice covered) salt lakes and salt oceans. Here the solubility of gases such as $CO_2$, CO, $CH_4$, $NH_3$, $NO_x$, $H_2S$, $SO_2$ in water and their corresponding ions is important and depends on pH, and the types of cations and anions present in solution.

The outgassed species in pyrovolcanism may undergo thermochemical reactions reaching equilibrium in vents and in plumes. On the other hand, the outgassed compounds of both, pyro- and cryovolcanism can undergo solar UV-driven photochemical (Moses et al. 2002a, b) and radiolytic (charged particle induced) reactions. The externally induced photochemical and radiolytic reactions also occur on exposed surface silicates, ices, salts, and organics which can contribute to tenuous atmospheres in addition to sputtering by micrometeoroid impacts. All these processes are important for long-term evolution of atmospheric and surface compositions (e.g., Huang et al. 2023).

Radiolysis is a major process affecting compositions because all moons considered here are exposed to strong particle radiation from their host planet's magnetospheres. These large and strong magnetospheres are due to strong dynamo action in the giant planet interiors and entrap charged particles from the solar wind and the planets themselves, but also receive ions from ejected plume material via neutral and ionized particle tori that in turn are fed by volcanically active moons along their orbits. Solar wind and planetary ions are mainly electrons, protons, with about 1-2 percent positively charged heavier elemental ions in about solar proportions. Plume material from Io provides S and O in the Jovian system; in the Saturnian system H and O are loaded into the magnetospheric flux by Enceladus. Ionized particles thus can be transported over wide distances from and to the planet and to its moons.

Jupiter's magnetosphere is 20-30 million kilometers in diameter, about 150 times wider than Jupiter itself and almost 15 times wider than the Sun. A larger fraction of the particles in the magnetosphere (and those implanted into the surfaces of the icy Galilean satellites) are derived from Io's plasma torus, which is fed via a neutral torus by Io's plumes rich in S-bearing compounds at an estimated rate of 1 ton/second (Thomas et al. 2004, see below). Saturn's magnetosphere is about 40 Saturn radii wide. It receives about 100 kg/s of H and O ions via charge exchange between neutrals in the Enceladus atmosphere and the corotating ions in Saturn's inner magnetosphere (Tokar et al. 2006). In turn, the source of atmospheric gases is water ice from cryovolcanism on Enceladus at a rate of about 600- 1000kg/s.

The Galilean satellites are all tidally locked. Except for Ganymede they do not have any significant intrinsic magnetic field and their surfaces therefore lack protection against charged particle radiation. Only Ganymede, the largest moon in the solar system, is shielded in the equatorial regions by an intrinsic magnetic field, which correlates with the presence of surface components such as salts and organics. Without the protection through a local magnetic field, these compounds would be destroyed by energetic charged particle radiation.

Photochemical and/or radiolytic decomposition (including sputtering) of gases or surface ices can change the overall chemical composition over time (e.g., Pollack and Witteborn 1980, Huang et al. 2023). Thus, the chemistry on the moons is not treatable as a "closed system", and there are external sources and sinks. The plumes erupting from Enceladus and the volcanos erupting on Io are dramatic examples showing at least two moons are not closed systems (e.g., Kargel 2006, Waite et al. 2006, 2017, Consolmagno 1979, Kieffer 1982).

Photolysis and radiolysis may lead to light gases and/or more volatile gases (e.g., $H_2$, CO, $N_2$) that could be lost from the moons, in particular the less massive moons over the age of the solar system (e.g., Lewis and Prinn 1984, Kargel 2006). A good example is loss of $H_2$, produced in net photochemical reactions such as

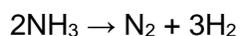
$2NH_3 \rightarrow N_2 + 3H_2$

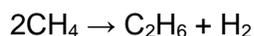
$2CH_4 \rightarrow C_2H_6 + H_2$

$2 H_2O \rightarrow O_2 + 2H_2$

$H_2S \rightarrow S(\text{elemental}) + H_2$.

These reactions can deplete the moons in hydrogen over time and may have played a role in Io's hydrogen depletion (along with thermally driven reactions releasing hydrogen). There are no definitive detections of hydrogen compounds in Io's atmosphere or surface, although Spencer and Schneider (1996) reported spectral features possibly due to $H_2O$ and $H_2S$ ices. Io is also apparently depleted in carbon and the absence of any CO or $CO_2$ in volcanic plumes places strict upper limits on Io's carbon inventory of C/S <$10^{-3}$ (Schaefer and Fegley 2005b). Feldman et al. (2004) reported a $C^{2+}/S^{2+}$ ratio of $3.7 \times 10^{-4}$ in the Io plasma torus, but as they discussed, the source of the carbon ions is uncertain and may come from the solar wind.

On the other hand, radiolysis of ices and possibly salts by energetic particles is responsible for $SO_2$ ice oxidation to $SO_3$ and sulfuric acid formation, and production of $H_2O_2$ and ozone from water which have been observed on several moons (see below). Photochemical and/or radiolytic driven decomposition of $CO_2$ ice via net reactions of the type $2CO_2$ (ice) $\rightarrow$ 2 CO (gas) + $O_2$ (gas) might successively deplete icy satellite surfaces in $CO_2$ ice and carbonates, or in the long run, in total carbon. Loss of oxygen as O or $O_2$ escape may also occur.

Loss considerations are generally important for the chemical evolution of the moons (e.g., Lewis and Prinn 1984). Cosmochemical models generally assume various volatiles (water ice, methane, ammonia, ice clathrates, ice hydrates and organics originally accreted to the moons. Some early models (e.g., Consolmagno 1979, 1981, Lewis 1982, Prinn and Fegley 1981) assume accretion of satellite – forming material in denser sub-nebulae present around the giant planets during their formation (e.g., Pollack and Reynolds 1974, Pollack et al. 1976) although more recent models suggest that satellite accretion occurred from a "gas starved" thin disk (e.g., Canup and Ward 2002). This difference is important because it affects the type of accreting low temperature species as well as the subsequent thermal and chemical processing during accretion of the moons.

The composition of any possible primordial atmospheres on the moons from evaporation of the original ices is not necessarily that what can be observed today. Erosion of atmospheres (plume ejecta, thermal and hydrodynamic escape, sputtering) can selectively remove gases. The composition of the outgassed species observed now is determined by what remains on the moons after 4+ Ga evolution: What is retained within the silicate lavas or aqueous source regions, what is released by hot and cold plumes. and what can be recycled on the moons or between moons and their host planet.

## Galilean Satellites

### Io

The Galilean satellite Io is a proto-type for a planetary object with abundant sulfur-driven chemistry. Since the 1970s, ground and space-based observations from mm, IR, optical to UV and spacecraft observations, mainly during the Voyager, Galileo, and JUNO missions, revealed several sulfur bearing gases and ices on Io. There is vast literature focusing on Io's sulfur chemistry and we refer to the book edited by Lopes et al. (2023) and the following reviews and collections: Kieffer (1982), Lewis (1982), Turtle and Thomas (2004), which is the special issue of Icarus on Io Galileo results, Lellouch (2005), Lellouch et al. (2015), Carlson et al. (2007), de Pater et al. (2023).

Figure Io1 gives a schematic overview of the sulfur cycle on Io, which is described below.

**Surface**. The burst in colors created by sulfur compounds on Io's surface was fully revealed by the Galileo spacecraft observations: Shades ranging from white, yellow, orange, to red with greenish and black tints and hues in-between (e.g., Geissler et al. 1999). White and gray moderate-to-coarse-grained $SO_2$ ice covers about 27% of Io, primarily in equatorial regions and localized at high latitudes. Yellow deposits make about 40% of the satellite's surface. Darker green and black deposits only cover a small fraction (about 1.4%) of the surface. While black color is a trademark of plastic sulfur made by rapid quenching of molten elemental sulfur, a shallow spectral absorption feature at 0.9 µm in most dark areas suggests the presence of Mg-rich silicates instead.

Geissler et al. (2004) describe surface changes on Io during the five years of the Galileo mission. About 27% of Io's surface changed in appearance testifying to its volcanic activity. Discoveries include newly recognized explosive volcanism centers and a few rare giant volcanic plumes ejecting dust in addition to gas. The outside areas of active volcanic structures ( = patera) became surrounded by colorful deposits, most likely of different sulfur modifications and allotropes from the magma flows. Compounds with polysulfur anions share the tendency to display similar wide color palettes as elemental sulfur. Impure sulfur with trace amounts of As and Se (Kargel et al 1999) also displays a wide color range. One or more of these compounds might be candidate species for the observed rings (see below).

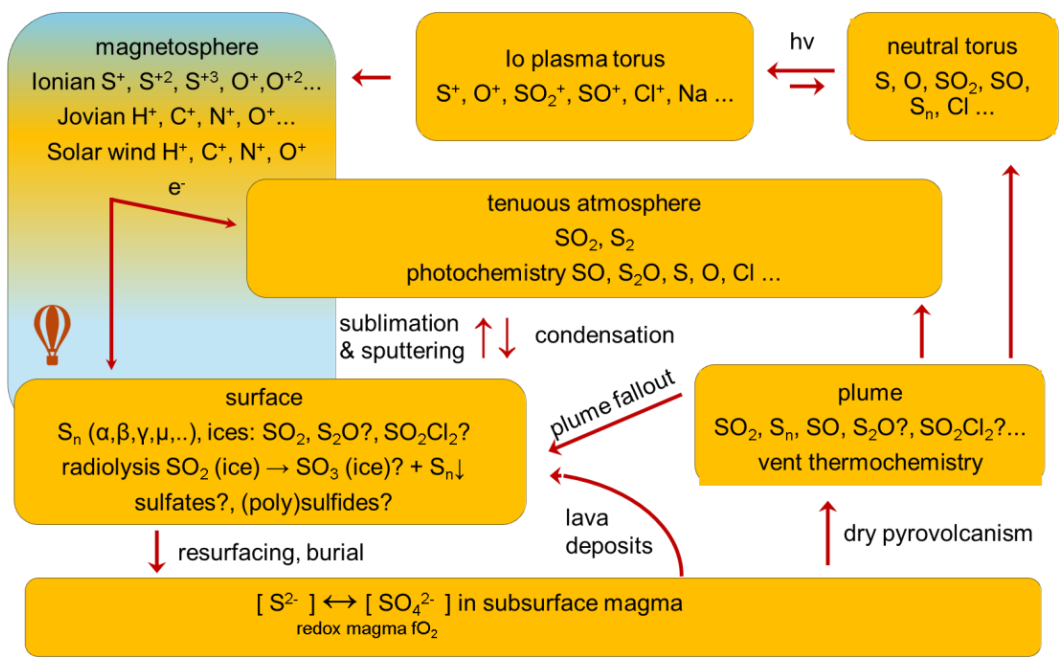

**Figure** 1. The sulfur cycle on Io showing sources and sinks, observed compounds, and inferred species that have been suggested to be present are marked with a "?".

**Volcanism**. Peale et al (1979) predicted volcanism on Io shortly before the discovery of the volcanic plumes. The occurrence of two different classes of volcanic eruptions was already apparent from Voyager 1 and 2 observations (McEwen and Soderblom 1983) and Geissler et al.

(2004) confirmed and extended the early results with Galileo data. In addition, Galileo data indicated regions with $SO_2$ seepage that was triggered by nearby volcanic activity. Eruptions of large plumes such as in Io's prominent volcano Pele are frequent and variable with short (minutes) outbursts of activity and some events ongoing over weeks to years (e.g., Radebaugh et al. 2004). These large plumes produced orange or red oval annuli that are stretched in north-south directions with maximum radii about 500-550 km. Using direct IR temperature measurements and the temperature-dependent color of elemental sulfur modifications, temperatures of around 650K could be associated with the regions of the large plumes (McEwen and Soderblom 1983). Galileo and Cassini observations indicated that the large region near Pele is ringed with bright hotspots that are relatively cool (< 800 K) but that the strongest thermal emissions reach $1500 \pm 80$ K (from the Cassini Imaging Science Subsystem, ISS, data) and $1605 \pm 220$ and $1420 \pm 100$ K (from a subset of Galileo Solid State Imaging, SSI, data) which are interpreted as regions of Pele's lava lake and upwelling and fountaining of basaltic lava at the lake's margins (see Radebaugh et al. 2004).

The more common smaller volcanoes such as the proto-type volcano Prometheus, have $SO_2$ rich plumes, have eruptions of longer durations (several years) and are responsible for most of the surface alterations. Estimated temperatures for the white and yellow (presumably sulfur) deposits are T<400 K. These deposits form near-circular structures about 150–200 km in radius. The smaller plumes also coat their surroundings with fine-grained $SO_2$-frost, a feature apparently not that common for larger plumes. The differences between plume classes is plausibly ascribed to the underlying causes driving the eruptions, which are closely related to the properties of sulfur (see below).

**Atmosphere and Volcanic Plumes**. Io's $SO_2$-rich atmosphere was discovered by Pearl et al (1979) and has been studied extensively (see below). Io's tenuous atmosphere ($\sim 10^{-9}$ bars) is largely made of $SO_2$, and it is spatially and temporally variable and not hydrostatically supported. Instead, it is supported by volcanic plume ejecta, outgassing, and $SO_2$ ice sublimation (e.g., Summers & Strobel 1996, de Pater et al. 2023). All gases ultimately originated from volcanic outgassing and are now entrained in the sulfur cycle on Io. The volcanic plumes contain mainly $SO_2$, SO, and $S_2$ (Spencer et al. 2000) with smaller amounts of $S_2O$ and S also expected (Zolotov and Fegley 1998a, b). Elemental sulfur gases (S, $S_2$) are largely restricted to Pele-type plumes, as are halides (NaCl, KCl) (Table 3). Volcanic emissions of $SO_2$ and of NaCl and KCl are not co-located. Monatomic gases (O, S, Na, K, Cl) and SO occur as volcanic gases and as photolysis products from molecular gases. Several reviews and summaries in papers cover observational results; see e.g., McGrath et al. (2004), Lellouch (2005), Moullet et al. (2010, 2013), and Redwing et al. (2022).

**Table 3.** Molecular Gases Observed on Io*

| Molecule | Volume Mixing Ratio | Where |
| --- | --- | --- |
| $SO_2$ | 1 | Global, patchy |
| $S_2$ | 0.08 – 0.33 | Plume |
| SO | 0.03 – 0.10 | Global |
| $S_2O$ | Not detected (yet?) | |
| NaCl | 0.003 – 0.013 | Plume |
| KCl | $4 \cdot 10^{-4}$ – 0.008 | Plume |

*See text and also summary in Moullett et al. 2013.

The plume chemical composition has been investigated using thermochemical (e.g., Zolotov and Fegley 1998a,b, 1999, 2000a, b, 2001, Schaefer and Fegley 2004, 2005a,b,c) and photochemical models (e.g., Moses et al. 2002a,b). Thermochemical models explore the expected gas speciation over bulk lava compositions that are constrained by cosmochemical and geochemical principles. Photochemical models explore how the equilibrium gas composition is altered by UV photons. Such theoretical calculations are quite useful to target and identify potential gas species in observational searches; such as predictions of NaCl, and KCl and subsequent discoveries. Other gases expected on the basis of thermochemical and photochemical calculations that are still to be found include allotropes of sulfur ($S_n$ with n =3-8), SCl, $S_2Cl$, $SO_2Cl_2$, $Na_2$, NaO, $NaO_2$, NaS, $NaS_2$, NaOS, $NaSO_2$, ClO, $Cl_2$, NaF, LiF, and LiCl (Schmitt and Rodriguez 2003, Fegley and Zolotov 2000a, Moses et al. 2002a, b, Schaefer and Fegley 2005).

Notably absent are water and $H_2S$, and nitrogen and carbon compounds; $H_2O$, $CO_2$, CO, $CH_4$, $H_2S$, COS, $CS_2$, $NH_3$, and $N_2O$ have been looked for but only detection limits exist. Chlorine is observed in Io's atmosphere (Feaga et al. 2004), but it is much less abundant than sulfur as expected from the solar, which is also the CI chondritic, Cl/S ratio of about 0.01 (Lodders 2020). As mentioned above, Io is dried up and must have had an extensive period of losses for volatile gases lighter than $SO_2$. Decomposition of water and loss of $H_2$ created an oxidizing environment, which is a source for converting sulfides to sulfur (if sulfides initially accreted), and then to sulfates during evolution (e.g., Lewis 1982). Atomic H (around 1 ppm by volume) was found in Io's atmosphere with HST observations, and protons constitute about 10% of the ion density of Io's torus (Strobel and Wolven 2001). This hydrogen is likely sourced via Jupiter's magnetosphere from Jupiter's upper atmosphere and from decomposing water ice on the other Galilean satellites. Overall, this small amount of H does not make H important for large scale chemical processes. Without abundant H for water and C as $CO_2$ for carbonates on Io, sulfur compounds are the major volatiles driving volcanism.

Although volcanic ejecta ($SO_2$ SO, $S_2$ and halides, and plume photolysis products such as S gas) will largely condense back on Io's surface, they also can be lost to space as plume ejecta can reach up to 500 km high. The loss of matter from Io into space is estimated as 1 ton/second (Thomas et al. 2004). The escaped gases travel as a structure of neutral gases ("neutral torus") along Io's orbit where $SO_2$, S, O, Na, K, and Cl have been observed (e.g., Thomas et al. 2004). Photochemical processes atomize and ionize matter, and ionization produces gases such as S+, O+, SO+, and $SO_2$+. This plasma becomes trapped in the rotating Jovian magnetic field forming a corotating ring structure called the Io plasma torus or just the Io torus. Molecular ions typically break up to atomic ions in the process. Ions from the torus can be scavenged onto paths along the magnetic field lines of Jupiter where they are also accelerated. These energetic charged particles populate the space of the vast magnetosphere of Jupiter. The S+ and O+ in the accelerated particle stream intercept Jupiter and also impact Io and the other Galilean satellites. Thus, interactions of ions in Io's torus with the Jovian magnetosphere provides a mechanism to transport abundant sulfur from Io to Jupiter as well as to Europa, Ganymede, and Callisto, where this "exogenic" sulfur can contribute to the formation of hydrated sulfuric acid upon radiolysis of water ice on the moons other than Io (see below).

Sulfur dioxide $SO_2$ gas and ice

Sulfur dioxide gas and ice is present on a global scale. The column densities of $SO_2$ gas are highest in equatorial regions and decrease with latitude (Jessup et al., 2004, 2007), and are somewhat higher on the non-Jupiter facing hemisphere. There are longitudinal variations with increase in $SO_2$ with proximity to volcanoes (e.g., Spencer et al. 2005, Moullet et al. 2013, Lellouch et al. 2015, dePater et al. 2023). Sulfur dioxide ice, $SO_2$ is widely distributed over the surface (Nash and Betts 1995, Doute et al. 2001). Early models already indicated that the atmosphere and its circulation are largely controlled by the evaporation-sublimation equilibria of $SO_2$ frost that is cold-trapped in the regolith, but that there is also a need for localized supply by surface venting and pyrovolcanic plumes (Matson and Nash 1983, Ingersoll et al. 1985, Ingersoll 1989, Thomas 1987) which is confirmed by recent observations.

There still is some debate about which process is dominant in supporting Io's atmosphere: sublimation and/or volcanic outgassing (see, e.g., Tsang et al. 2016, dePater et al 2023). Both processes are needed to maintain a "permanent" $SO_2$ atmosphere. There are direct losses of $SO_2$ from volcanic ejecta into Io's neutral torus. Other losses of atmospheric and surface $SO_2$ ice to the magnetosphere are by sputtering and radiolysis with ions from the corotating Io plasma torus to which they are also lost. Cold-trapping of $SO_2$ at the polar regions could deplete the $SO_2$ atmosphere at mid latitudes if volcanic outgassing were not compensating any losses.

A characteristic near-surface (kinetic) gas temperature is difficult to obtain and remains somewhat uncertain, which affects the quantitative assessment of the $SO_2$ concentrations from observations (further omitting localized temperature increases due to volcanism). This situation is described in Lellouch et al. (2015). Considering only solar heating at heliocentric distance of 5.07 AU gives Io's peak dayside surface temperature as 116.2 K (using a Bond albedo of 0.55 and thermal inertia of 400 W m$^{-2}$ s$^{-1/2}$ K$^{-1}$), the noon temperature is 115.0K, at -40° and +80° longitude from noon the temperatures are 111.0 and 113.6 K respectively. The photopolarimeter/radiometer (PPR) on GALILEO determined 130K for the dayside from the intensity and polarization of reflected sunlight (Rathbun et al. 2004).

The derived gas kinetic temperatures (atmospheric temperatures) from spectral modelling seem to have some unexplained dependence on spectral wavelengths. Derived temperatures for Io are e.g., 150-220K with a mean of 170K (4-µm Lellouch et al. 2015), less than 140 to 150K on the anti-Jupiter hemisphere (19-µm, Spencer et al. 2005), and 115K (19-µm,Tsang et al. 2016). DePater et al. (2020) reported a range of 220K -320 K for kinetic temperatures near individual volcano locations during eclipse and in sunlight from disk integrated ALMA spectra. This temperature range seems extremely high for a "global average" temperature.

Estimates for characteristic surface pressure on Io critically depend on the surface and gas temperatures. One order of magnitude estimate for the total pressure is about 1- 10 nbar of mainly $SO_2$ around 115K comes from a frequently used vapor pressure equation for $SO_2$ gas over $SO_2$ ice (T≤197.7 K):

$$logP_{SO_2}(bar) = 8.1818 - \frac{1958.67}{T}$$

This equation is generally cited in the literature as coming from an obscure report by D.D. Wagman at NBS(NIST) in 1979. A literature survey shows actual measurements of the vapor pressure over $SO_2$ solid are limited to temperatures from 170 – 197.7 K, which is the triple point of $SO_2$ according to the NIST Chemistry WebBook. The linear fit values are thus generally extrapolated quite a bit for application to Io's surface temperatures. A linear fit equation assumes a constant enthalpy of sublimation and no difference in the heat capacity of $SO_2$ gas and ice. Neither assumption is correct but may be useful over limited temperature ranges (see pp. 250-256 in Fegley 2013). Instead, we derived the equation below:

$$lnP_{SO_2}(bar) = 27.140 - \frac{4{,}594.6}{T} - 1.51lnT$$

This incorporates an average heat capacity difference between gas and ice and gives a $SO_2$ vapor pressure of 10 nanobar at 120 K. For reference the vapor pressure data of Honig and Hook (1960) give 13 nanobar ($10^{-5}$ mm Hg) at 119.5 K and Wagman's linear fit equation gives 7.2 nanobar at 120 K. In any case, the steep exponential increase of vapor pressure with temperature gives orders of magnitudes in pressure increases, e.g., the $SO_2$ vapor pressure over $SO_2$ ice is 180 nanobar at 130K and 0.48 millibar at 170 K (from the 3-term equation).

The temperature range of 220-320K near individual volcanoes observed by de Pater et al. (2020) is above the triple point of $SO_2$ (197.7 K). The vapor pressure of $SO_2$ over liquid $SO_2$, taken from Giauque and Stephenson (1938) from the triple point to the one atmosphere boiling point of 263.5 K is:

$$logP_{SO_2}(bar) = 10.2003 - \frac{1867.52}{T} - 0.015865T + 1.5574 \times 10^{-5}T^2$$

At 220 K, this gives 94 mbar. If these temperatures apply to gas above the surface, rapid cooling of such atmospheric $SO_2$ gas might provide liquid $SO_2$ fog and mist instead of $SO_2$ snow. As the global average surface temperature remains unclear, not much can be said about the near surface total pressures; however, comparisons of relative variations from observed column densities are possible.

The vapor pressure supported $SO_2$ atmosphere reveals itself during Io's eclipses. Tsang et al. (2016) observed two eclipses at 19-μm with GEMINI and found that while surface temperatures dropped quickly from 127K to 105K after onset of the eclipse, the column density of $SO_2$ dropped rapidly by a factor of 5 (±2) from pre-eclipse to mid eclipse after 40 minutes. Using the 3-term vapor pressure equation for $SO_2$ ice above, the change in vapor pressure from 127K (79 nbar) to 105K (0.054 nbar) is about a factor of 1500, far higher than the observed drop of five in column densities. For reference, the Wagman linear fit vapor pressure equation gives a factor of about 1700. This large discrepancy shows that the approximation of surface- atmosphere equilibrium (implicit in the use of the vapor pressure equation) is an oversimplification. Other factors must be important for the $SO_2$ column density.

Using ALMA, de Pater et al. 2020 found that during ingress the drop in solar heating is accompanied with an exponential decline in $SO_2$ flux and drop in column densities by a factor of 2-3. On egress, the $SO_2$ atmosphere rebuilds linearly within 10 minutes. However, they also found some low-level $SO_2$ emissions during the eclipse which they ascribe to "stealth volcanism". These could be volcanic vents without prominent large scale ejecta plumes. DePater et al. (2020) also found an area (Ulgen Patera) where SO and $SO_2$ seem to be absent yet KCl gas is present suggesting different compositions for different magma chambers.

### Disulfur, $S_2$, and condensed elemental sulfur modifications

Gaseous $S_2$ was detected with the Hubble Space Telescope (HST) in Pele's plume where it has time-variable mixing ratios ranging from $S_2/SO_2$ = 8 − 30%, accompanied by atomic S (Spencer et al. 2000, McGrath et al. 2000, Jessup et al. 2004, 2007). The $S_2/SO_2$ ratios are exceptionally high for the Pele vent. Observation of $S_2$ at other locations are not that frequent, and $S_2$ is not detected in all volcanic plumes (Jessup et al 2007). Disulfur is more refractory than $SO_2$ and condenses in the vicinity of volcanoes.

Detailed thermochemical computations of the sulfur-oxygen gas chemistry by Zolotov and Fegley (1998a, b, 1999, 2000a, 2001) can be used to narrow the temperatures, pressures, total S/O ratios and redox conditions in Io's volcanic vents from the observed relative abundance of $SO_2$, $S_2$, SO, and S. At equilibrium, the gas chemistry of all sulfur and oxygen gases is coupled and individual equilibria have different dependencies on the parameters of temperature, total pressure, and oxygen fugacity so that ratios of individual gas pairs can be used to constrain these parameters. For example, the gas phase equilibrium

$$SO_2 + S = 2\ SO\ (g)$$

is independent of total pressure ($P_T$) as shown by Dalton's Law and the equilibrium constant ($K_{eq}$) expression:

$$P_i = X_i P_T \text{ Dalton's Law}$$

$$K_{eq} = \frac{P_{SO}^2}{P_{SO_2} P_S} = \frac{X_{SO}^2}{X_{SO_2} X_S}$$

The equilibrium constant is only dependent on temperature, and can be computed using the standard Gibbs energy of reaction (from compiled thermodynamic data) via the equation:

$$\Delta G° = -RT ln K_{eq}$$

The observed abundances of $SO_2$, SO, and S in a volcanic plume give an independent value of $K_{eq}$ and the temperature at which the two values are equal constrains the gas temperatures to 1440±150 K (Zolotov and Fegley 2000a). Such high temperatures are broadly consistent with thermal observations of volcanism on Io (dePater et al 2021).

The oxygen fugacity $fO_2$ at the derived vent temperature can be constrained by any of several different pressure independent reactions (Zolotov and Fegley 2000a):

$$2\ SO_2 = 2\ SO + O_2$$

$$2\ SO = 2\ S + O_2$$

$$SO_2 = S + O_2$$

$$3\ SO_2 = 2\ SO + S + 2\ O_2$$

For example, the $fO_2$ for thermal decomposition of $SO_2$ to SO is given by the equations:

$$K_{eq} = \frac{X_{SO}^2 f_{O_2}}{X_{SO_2}^2} = exp\left(-\frac{\Delta G°}{RT}\right)$$

$$f_{O_2} = K_{eq} \frac{X_{SO_2}^2}{X_{SO}^2}$$

The computed oxygen fugacities are about 3.3 log units below that of the nickel-nickel oxide buffer (NNO) used in petrology:

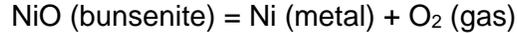

$$NiO \text{ (bunsenite)} = Ni \text{ (metal)} + O_2 \text{ (gas)}$$

As shown in Figure 2 of Zolotov and Fegley (2000a), the $fO_2$ from the NNO buffer reaction is temperature dependent, as are $fO_2$ values from other buffer reactions, e.g., the sodium sulfide – sodium sulfate buffer, which may be relevant to the more oxidized Loki plume (Zolotov and Fegley 1999). Thus, $fO_2$ is more conveniently expressed relative to that of an $fO_2$ buffer rather than as absolute values.

The total pressure of the erupting volcanic vents depends on the bulk O/S abundance ratio. The observed gases are dominated by $SO_2$ and $S_2$ and the comparison of observed outgassed $SO_2/S_2$ with gas equilibria suggests a bulk O/S ratio around unity, corresponding to a 1:2 $S_2/SO_2$ mixture. Combining the results for temperatures, oxygen fugacities and O/S ratio gives estimates for total pressures of the volcanic vent at Pele of $10^{-4.7}$ to $10^{-5.4}$ bar (Zolotov and Fegley 2000a). The vent pressure calculated for Pele is orders of magnitude larger than the nominal average surface pressure of about 10 nanobar ($10^{-8}$ bar).

Precipitation of sulfur from volcanic plumes rich in $S_2$ and from sulfur saturated lavas are sources for elemental sulfur surface deposits on Io. Condensed sulfur has several allotropes, two major liquid structures (chains and ring structures) and polymerizes into "polysulfur". The phase diagram is quite complex. Different elemental sulfur allotropes contribute to Io's colorful surface appearance. Allotropes of elemental sulfur $S_n$ (with n = 2-8) and polysulfides were already suspected to be present on Io by Wamsteker et al. (1974) who used reflection spectra of the Galilean satellites ratioed to that of Saturn's water ice rings to identify surface materials. Wamsteker et al. (1974) noted the resemblance of Io's differential reflection spectrum to that of cold sulfur (at 113 K) measured by Sill (1973). It has been suggested that the reddish rings around Pele are possibly $S_3$ and $S_4$ modifications, but other explanations are also possible. The formation conditions often determine which sulfur modification forms, and the stability of allotropes is sensitive to external variables such as photon and energetic particle exposure. Upon temperature changes, some modifications only slowly convert from their metastable states into the thermodynamically stable modification, hence associated changes in color appear gradually. Kargel et al. (1999) list four possibilities explored by numerous authors for the reddish colored rings: (1) quenching of molten sulfur, (2) coloration of $SO_2$ or S by $S_2O$, (3) coloration of elemental sulfur by radiation, and (4) impurities included in sulfur. We refer the reader to Kargel et al. (1999) for more discussion of this colorful topic.

Sulfur monoxide, SO, Disulfur oxide, $S_2O$, and sulfur trioxide $SO_3$,

Sulfur monoxide was detected by Lellouch et al. (1996) and occurs globally at a mixing ratio of 3–10% of $SO_2$. Typically, the sulfur oxides SO and $SO_2$ occur together as their chemistry is linked, and SO is often more concentrated towards volcanic regions (e.g., de Pater et al. 2002, 2007; Moullet et al. 2010, 2013). In plume gases they are often observed together with monatomic S by UV spectroscopy (McGrath et al. 2000, Jessup et al. 2004) and with $S_2$ (Spencer et al. 2000; Jessup et al. 2007). Sulfur monoxide gas from the plume is also fed into Io's exosphere, where it is observed as $SO^+$ together with $SO_2^+$ (Russell and Kivelson 2000).

Like $SO_2$, the observed SO column densities and distributions require at least two different sources of SO (Moullet et al. 2010, 2013). The production of SO in volcanic gas equilibria (Zolotov and Fegley 1998a) occurs highly localized and does not supply observed SO concentrations in areas without or with low volcanic activity. However, photolysis of $SO_2$ can occur in any regions exposed to solar UV photons. The photolysis also leads to S, O, $O_2$, and $S_2$ (Summers & Strobel 1996). Especially $O_2$ has not yet been detected but would be an important tracer for this process. Another product of photolysis should be disulfur dioxide, $S_2O_2$, which is the dimer of SO.

The photochemical mechanism of SO production from $SO_2$ gas is inferred from the eclipse studies of Io. For example, de Pater et al. (2020) observed that the decrease in SO during eclipse ingress and increase during egress followed that of $SO_2$ but was more gradual. This is interpreted as the absence of photons stopping production of SO from $SO_2(g)$. The other reason for the decrease in SO is of course the decrease in the target gas $SO_2$ through condensation. Both the decreased photon flux and the condensation of $SO_2$ decrease the SO abundance.

It was thought that SO could condense (e.g., Schenk 1933) but unlike $SO_2$, SO does not condense. It is a highly reactive biradical with two unpaired electrons and is unstable as a liquid or solid unlike $O_2$, which has the same electronic ground state (Schenk and Steudel 1965). Usually SO disproportionates via the net reaction

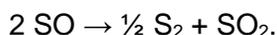

$$2\ SO \rightarrow \tfrac{1}{2}\ S_2 + SO_2.$$

Generally, SO(g) is very unstable at low temperatures in denser gases which may also provide a constraint on the total atmospheric pressure (see section on $SO_2$ above).

Disulfur oxide ($S_2O$, also called sulfur suboxide) is a gas expected in Io's atmosphere but searches have not been fruitful (Moullet et al. 2010). Both suboxides, SO and $S_2O$, share chemical similarities. Both are minor gases in equilibrium with $SO_2$ and $S_2$ in Ionian volcanic gas (Zolotov and Fegley 1998a,b). Electric discharge and photolysis of $SO_2$ produces SO and $S_2O$, and SO disproportionates to $S_2O$ via the net reaction $3\ SO \rightarrow S_2O + SO_2$. Pure $S_2O$, a cherry-red condensate at liquid air temperatures, has been known in the chemical literature for some time. Meschi and Myers (1956) showed that Schenk's sulfur monoxide (Schenk 1933) was actually a mixture of mainly $S_2O$ with some $SO_2$. Upon warming $S_2O$ disproportionates into $SO_2$ and yellow plastic polysulfane oxides containing chains of $[S_3O]_n$ and $[S_4O]_n$ that polymerize via the net reactions

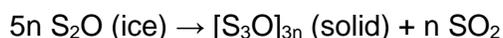

$$5n\ S_2O\ (ice) \rightarrow [S_3O]_{3n}\ (solid) + n\ SO_2$$

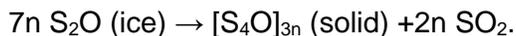

$$7n\ S_2O\ (ice) \rightarrow [S_4O]_{3n}\ (solid) + 2n\ SO_2.$$

Increasing temperature causes decomposition of these polysulfane oxides to $SO_2$ and polysulfides $S_n$ that convert to orthorhombic $S_8$ by about 373K. Steudel (2003b) describes chemistry of $S_nO$ ($n \geq 2$ in detail).

Condensation of $S_2O$ has been considered to explain the reddish coloration of deposits around Io's volcano Pele (cf Kargel et al. 1999). This was tested by Baklouti et al. (2008) with condensation experiments where $SO_2$ was converted to $S_2O$ polymers by electric discharge under conditions applicable to Io. They discuss the possible identification of polysulfuroxide in NIMS/Galileo infrared spectra using their lab spectra for identifications, but firm conclusions are difficult. Baklouti et al. (2008) concluded that a mixture of orthorhombic $S_8$ and $SO_2$ gave the best match to Io's reflectance spectrum in the 330 – 520 nm range but that $S_2O$ polymers may be important in some volcanic regions.

As mentioned earlier, Khanna et al. (1995) reported the possible identification of $SO_3$ ice on Io's surface. Pearl et al (1979) discovered $SO_2$ in the Loki volcanic plume and gave an upper limit for SO$_3$, which is a mole fraction X(SO$_3$) <2×10$^{-4}$ (Zolotov and Fegley 1999). The actual abundance of SO$_3$ in Ionian volcanic gases is plausibly orders of magnitude lower (<10$^{-7}$) as detailed by Zolotov and Fegley (1999, 2000a). Sulfur trioxide is absent from Ionian volcanic gases for all practical purposes. The source of SO$_3$ ice on Io's surface may be proton irradiation of SO$_2$ ice as found by Moore (1984) and Strazzulla et al. (2009). Mifsud et al. (2021) discuss the possible reaction mechanisms in more detail.

## Halogenated sulfur compounds

The presence of NaCl and KCl in plumes is a possible source of Cl for chlorinated gases and ices on the surface. If volcanic plumes contain more chlorine than Na + K, which may occur in low temperature plumes, then Cl is available to form Cl, Cl$_2$, S$_2$Cl, SCl$_2$, SOCl$_2$, S$_2$Cl$_2$, and SO$_2$Cl$_2$ (Fegley and Zolotov 2000). Only monatomic Cl has been detected to date (Feaga et al. 2004). The chlorine sulfur gases are also produced in photochemical reactions of volcanic plume gases with dichlorosulfane, SCl$_2$, being a major product (Moses et al. 2000). Thionyl chloride, SOCl$_2$ only becomes important under conditions when Na and K are depleted in the gas (e.g., Na/Cl and K/Cl ratios far below stoichiometric ratios in the halides). If present photolysis of SOCl$_2$ should also yield the ClSO radical. The survival and potential (co)condensation into SO$_2$ ice or in sulfur deposits requires a UV-shielded environment.

Another formation mechanism of halogenated sulfur compounds is chlorination of SO$_2$ ice or elemental sulfur. Condensed NaCl, S and SO$_2$ can coexist on the surface. Sputtering and energetic particle radiation may induce reactions. Singly and doubly ionized Cl is present in the Io torus (Küppers and Schneider 2000, Feldmann et al. 2001) and is a reactive species in the magnetospheric charged particle stream onto Io. Thus, some chlorinated ices and liquids are expected which can also contribute to Io's color palette, e.g., liquid SCl$_2$ (mp. 151.2 K) is dark yellow brown to cherry red and was proposed by Schmitt and Rodriguez 2003 as an alternative explanation for the reddish coloration of some volcanic deposits. Similarly, chloropolysulfanes can be expected. Note the oxidation state of S$^{2+}$ in SCl$_2$, thus reduction of S$^{4+}$O$_2$ is involved. This is of interest as other reactions tend to oxidize, rather than reduce sulfur.

The infrared spectrum and vapor pressure of sulfuryl chloride Cl$_2$SO$_2$ was investigated by Schmitt and Rodriguez (2003) who found that condensed sulfuryl chloride Cl$_2$SO$_2$, and possibly chlorosulfonyl, ClSO$_2$, in SO$_2$ ice are candidates for the absorber(s) of the 3.92 mm band locally present in NIMS/Galileo spectra of the reddish deposits south of Io's Marduk volcanic center. They favor a heterogeneous reaction of Cl atoms on SO$_2$ ice condensing on plume particles or at Io's surface; here the oxidation state of S$^{4+}$ in SO$_2$ switches to S$^{6+}$ in sulfuryl chloride.

## Other polysulfides

Elemental, oxide, and chlorinated polysulfides are mentioned in the previous sections, and here we note other polysulfide candidates and tentative detections on Io's surface.

The classic polysulfanes H-S$_n$-H with n = 2-8 and higher homologues of H$_2$S, also called hydrogen polysulfides, are not expected in the SO$_2$ rich environment on Io, nor is H$_2$S ice. It was thought that H$_2$S is present (Nash and Howell 1989, Salama et al. 1990) as these ices seem to match the 3.92 micron absorption feature, but this match is non-unique (Schmitt and Rodriguez 2003). If present, H$_2$S decomposition accompanied by H$_2$ loss would leave polysulfanes at low temperatures: n H$_2$S = H$_2$S$_n$ (polysulfane) + H$_2$↑.

Reactions of polysulfur with implanted protons may also produce occasional $H_2S_n$. The proton irradiation experiments of H+ onto $SO_2$ ice by Strazzulla et al. (2009) produced $SO_3$, and polymers of $S_n$, and $O_3$; they did not find $H_2SO_3$ in their experiments and rule out $H_2SO_3$ for dry Io. They did not observe H-S formation but irradiation plus warming induced the formation of a sulfurous residue that showed S-H bonds, probably including $H-S_n-H$ polysulfanes from the $S_n$ resulting from $SO_2$ ice disproportionation.

Alkali polysulfides may occur in Io's magmas depending on oxygen fugacities and sulfur species dissolved in the magma (see below). Similarly, the plume chemistry of S-O-Na-Cl predicts the condensation $Na_2S$ at higher temperatures and lower O/S ratios whereas $Na_2SO_4$ appears at lower temperatures and higher O/S ratios, in both cases accompanied by copious NaCl and KCl (Zolotov and Fegley 2000a). We speculate that surface deposits of alkali polysulfides and alkali halides may show interesting phase exchange equilibria with thermochromic effects. Some conditions allow co-condensation of $Na_2SO_4$ and $Na_2S$, analog to the coexistence of sulfides and sulfates dissolved in magma as a function of oxygen fugacity (see below).

The Na and K polysulfides are yellow to orange-red in color (e.g., Steudel 2003a). Their phase diagrams are complex, and these thermochromic polysulfides could contribute to temperature-dependent reddish surface hues on Io's surface. However, the alkali polysulfides would be difficult to distinguish from elemental sulfur as in both the sulfur chains determine the major spectral features and IR active functional groups (S-H, S=O, S-Cl) are not present.

In solution, polysulfide anions, $S_n^{2-}$, are in equilibrium with their mono radical anions, $S_n^{2-} = 2S_{n/2}^{-}$ which are often responsible for the strong colorations in polysulfide compounds (e.g., Steudel 2003a). If produced under influence of radiation in surface deposits, polysulfide mono- and bi radical anions may also contribute to Io's colors. Polysulfides are sensitive to oxidation which adds the possibility of thiosulfates, $S_2O_3^{2-}$, into the mix of possible surface coatings on Io.

## Sulfur in Io's magma

Sulfur and its oxides are the major volatiles on Io and are responsible for fostering Io's volcanic activity. There are no unambiguous detections of hydrogen, carbon, or their compounds on Io and these elements seem to have been lost from Io over time (Consolmagno 1979, 1981, Lewis 1982, Zolotov and Fegley 2000b, Schaefer and Fegley 2005b). Any hydrogen brought in from solar wind (major element there) or transported from Jupiter via the magnetosphere appears to be negligible or surface limited. The absence of water and $CO_2$ is in stark contrast to terrestrial volcanism where water and $CO_2$ are usually the two major volatiles.

Compounds of sulfur and the halogens salts are more refractory than water ice. Several models consider that Io (and other satellites) formed from material akin to carbonaceous chondrites, and taking the CI-chondrites suggests initial mass ratios of S:Cl:F = 5.36:0.072:0.009. Thus, sulfur compounds are the prime candidates to replace water and $CO_2$ as volatile constituents in magmas and lavas. Dissolved sulfur dioxide, $S_2$, sulfates, sulfides, and other S-compounds in magmas would cause the same effects as water in magmas on Earth: lowering the melting point of melts, affecting viscosity and thus influencing magma eruption dynamics.

The presence of $SO_2$ and $S_2$ gas in plumes leads to the question in what form sulfur is dissolved or associated with in the magmas, which depends on the oxygen fugacity. The results by Zolotov and Fegley (1999, 2000a; see also above) for gas chemistry in Io's volcanic gases indicated oxygen fugacities, $fO_2$ up to three logarithmic units lower than the nickel – nickel oxide buffer (NNO-3.3). This oxygen fugacity should also apply to the magma from which the gases were exsolved and/or has equilibrated with before and during eruption. This range in oxygen fugacities is comparable to that found in terrestrial basaltic magmas and their associated volcanic gases. This similarity makes it plausible that silicate volcanism on Io is driven by sulfur.

Sulfur can be dissolved in silicate melts as both $SO_4^{2-}$ and $S^{2-}$, which depends mainly on the oxidation state (oxygen fugacity) of the melt and cations present; total pressure effects are secondary here. Several experimental studies show that sulfur has a minimum solubility at about the NNO buffer (e.g., Haughton et al. 1974, Katsura and Nagashima 1974, Wallace and Carmichael 1992, Carroll and Webster 1994). The reason for this minimum is that the oxygen fugacity has to be high enough to stabilize sulfate or low enough to allow sulfide in solution.

Oxidation of neutral sulfur $S^o$ or $S^{4+}$ in $SO_2$ to $S^{6+}$ to sulfate increases sulfate with CaO (or also $Na_2O$, MgO) in melts:

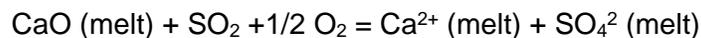

$$CaO\ (melt) + SO_2 + 1/2\ O_2 = Ca^{2+}\ (melt) + SO_4^2\ (melt)$$

or

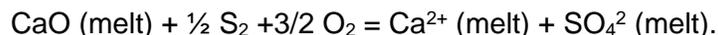

$$CaO\ (melt) + ½\ S_2 + 3/2\ O_2 = Ca^{2+}\ (melt) + SO_4^2\ (melt).$$

At high $SO_2$ and high oxygen fugacities, melts can become saturated in sulfate and exsolve anhydrite, $CaSO_4$. Experiments with dry $SO_2$ acting on basalt as function of oxygen fugacity further showed production of $MgSO_4$ and Na-sulfates but not Fe-sulfates (e.g., Renglii et al. 2019). The presence of such sulfates on Io's surface is surmised but not firmly established from spectroscopy. Suggested sulfates include Na and K-sulfates, and the suite of Mg-sulfates ($MgSO_4 \cdot nH_2O$ with different crystal water content of n = 0-6). As water is not known in Io's volcanic plumes, the presence of hydrated sulfates in lava flows at the surface is unlikely; however, one could speculate that hygroscopic $MgSO_4$ may have acquired hydration from water that possibly formed from exogenic protons (e.g., see Kargel 1992).

In this context it is noteworthy that radiolysis of water-free $SO_2$ ice leads to disproportionation reactions such as $3\ SO_2 = 2\ SO_3 + S$ (e.g., Moore 1984; see also above). This brings sulfate production a step closer if $SO_3$ bearing ice is buried by resurfacing and introduced into magma. The radiolysis is also another possible source for the sulfur allotropes that are causing the surface colorations and spectral darkening on Io.

The reduction of $SO_2$ and $S_2$ to $S^{2-}$ is favored on the low $fO_2$ side of the sulfur capacity minimum:

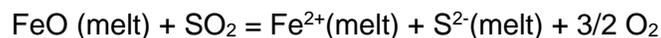

$$FeO\ (melt) + SO_2 = Fe^{2+}(melt) + S^{2-}(melt) + 3/2\ O_2$$

Or

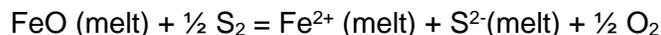

$$FeO\ (melt) + ½\ S_2 = Fe^{2+}\ (melt) + S^{2-}(melt) + ½\ O_2$$

An excess of sulfur under reduced conditions leads to iron sulfide formation, but alkali and Ca sulfide, CaS, and polysulfides may also occur, again depending on oxygen fugacity, total sulfur in the system, and cation availability. Calcium sulfide solubility in and precipitation from melts is well known and studied by metallurgists (e.g., Fincham and Richardson 1954, Sharma and Richardson 1962, Turkdogan and Darken 1961). Magnesium sulfide is much less soluble than CaS under the same conditions (Abraham and Richardson 1960). There seems to be no evidence for the presence of iron sulfides such as pyrrhotite ($Fe_{1-x}S$, $Fe_7S_8$), or troilite (FeS) on Io. The occurrence of pyrite or marcasite ($FeS_2$) has been suggested by Kargel et al. (1999) but is also not conclusively identified yet.

All sulfur dissolution reactions above involve $O_2$ and one open question is what is controlling the oxygen fugacity in Io's magmas. The oxygen fugacities inferred are around NNO-3 from the plume gas compositions (Zolotov and Fegley 1999, 2000). Possible melt equilibria must involve an abundant element which leaves the redox pair of $Fe^{2+}$ and $Fe^{3+}$ as a first plausible choice. However, melt equilibria such as

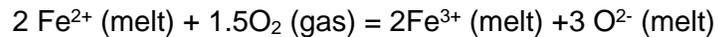

$$2\ Fe^{2+}\ (melt) + 1.5O_2\ (gas) = 2Fe^{3+}\ (melt) + 3\ O^{2-}\ (melt)$$

are not univariant and are not buffers because the activities of $Fe^{2+}$ and $Fe^{3+}$ are not uniquely fixed within the melt (While the absolute concentration of Fe may not be variable in the melt, the distribution between $Fe^{2+}$ and $Fe^{3+}$ is). A potential "semi buffer" involving $Fe^{2+}$ in a melt saturated with magnetite such as

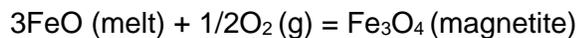

$$3FeO\ (melt) + 1/2O_2\ (g) = Fe_3O_4\ (magnetite)$$

is also not univariant. If Fe sulfides are coexisting with the magmas, the redox equilibria involving iron become even more complex. At present, this is an area where more modelling and experimental work is required.

## Icy Galilean Satellites

We now come to the other Galilean satellites, which, in stark contrast to Io, contain larger fractions of water ice and in part organics and ices of other highly volatile compounds. They are all less dense than Io as illustrated in Figure 2 for a collection of moons of the gas giant planets. Among planet and moon building materials, water ice, organics, and other icy compounds have lower densities than "rocky" (i.e., metal, sulfide, silicate) planetary materials.

The proportions of their initially accreted rocky and icy components (ice:rock ratios), and the nature (composition, redox states) of these components control the outcome of their large-scale differentiation during and after accretion. Among the Galilean moons the pronounced density drop from Io and Europa to Ganymede and Callisto heralds a shift away from the internal structures of largely metallic core and silicate mantle towards metal-free, hydrous silicate cores with considerable size icy mantles and icy crusts. For sulfur, this means a shift from being a possibly significant constituent of a metallic core and actor during silicate magmatic mantle processes to becoming trapped as sulfide or sulfate anion in ice, aqueous brines, evaporite salts, and salty oceans. This is more important than often realized – it takes salt to make and sustain liquid subsurface oceans that are now known to be present on several moons, and sulfur can have a large role in that (Kargel and Consolmagno 1996). While halides are important and NaCl dominates the terrestrial ocean salts, sulfur is generally much more abundant than chlorine in the makeup of an entire planet or moon. Subsurface oceans are quite plausible on Europa, Ganymede, but also suspected on Callisto, and we mention here the Saturnian moons Enceladus and Titan, that also harbor oceans. All these moons have larger diameters (see Figure 2), and Dione, one of the next larger moons and neighbor of Enceladus, also joined the ocean club. Neptune's moon Triton is the seventh largest moon and plots together with other ocean bearers in Figure 2, and yes, it also shows evidence for a subsurface ocean.

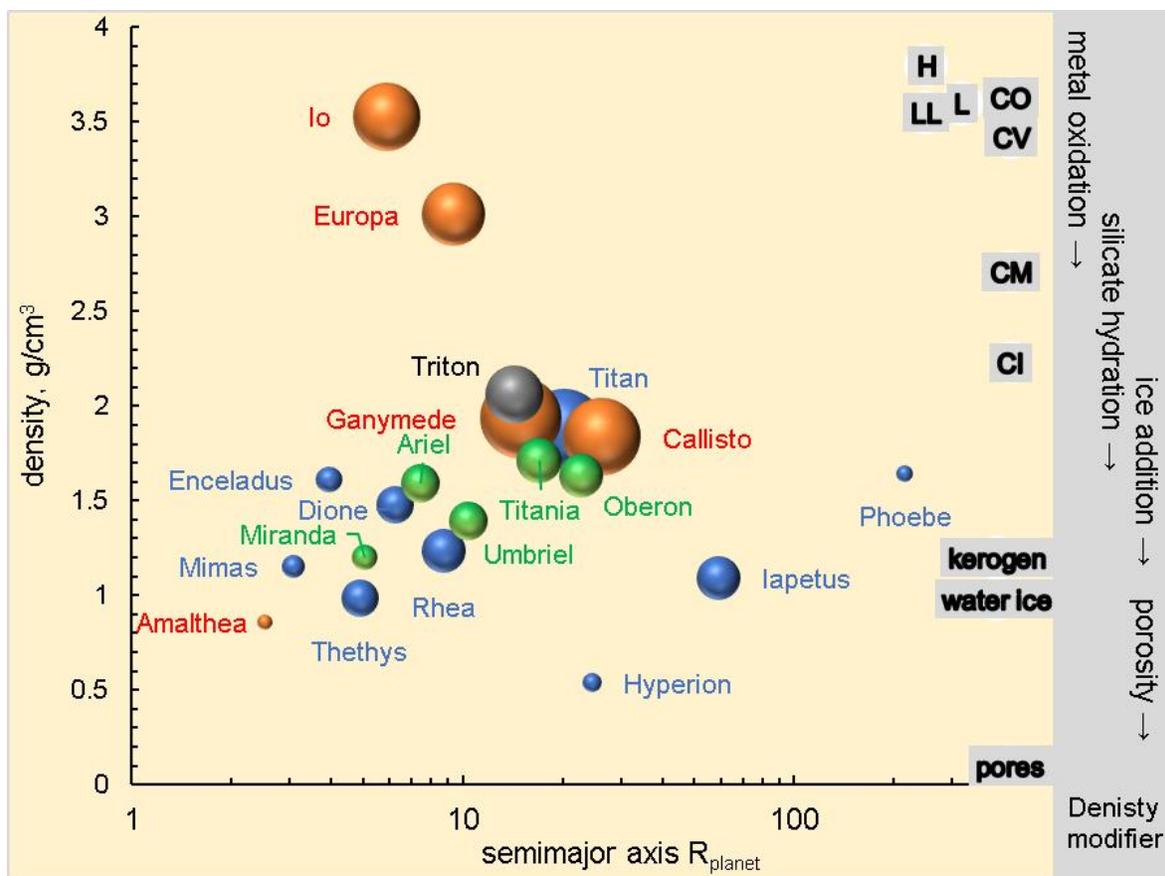

**Figure 2**. The densities of major moons of the gas giant planets as function of radial distance from the planetary hosts in respective planetary radii. The symbol size indicates the relative diameters of the moons. Jovian = orange. Saturnian = blue. Uranian = green. Neptunian = grey. On the right densities of chondritic meteorites, kerogens, and water ice are indicated and how densities are affected by different processes that act as density modifiers.

What happens beneath, in, and on the icy surface shells is largely determined by the moons' orbital distance from Jupiter, their tidal interactions that provide internal heating for driving chemical reactions and geological actions, and time.

The geophysical and geological observations of the Jovian satellites Europa, Ganymede, and Callisto, and the Saturnian satellites Enceladus are covered in the review by Soderlund et al. 2020 and we focus on the geochemical aspects concerning the sulfur chemistry here. The literature is extensive and we refer the readers to the following papers by Pappalardo et al. (2009 on Europa), and Volwark et al. (2024 on Ganymede).

Before going into the sulfur chemistry, we briefly describe a few properties of Europa, Ganymede, and Callisto in the next subsections. Table 4 summarizes some of their identified surface compounds and molecular gases in their very tenuous atmospheres which is for reference in the discussion below.

**Table 4.** Surface Compounds and Atmospheric Molecules Observed on Europa, Ganymede, and Callisto

| Compound | Europa | Ganymede | Callisto |
|---|---|---|---|
| Water ice | Y | Y | Y |
| $CO_2$ in water ice, $CO_2$ atmosphere" | Y | Y | Y |
| | | | Y |
| $SO_2$ in water ice $SO_2$ atmosphere | Y | Y | ? |
| $O_2$ in water ice | Y | Y | Y |
| $O_2$ atmosphere | Y | Y | Y |
| $O_3$ in water ice | Y | Y | Y |
| $O_3$ atmosphere | ? | Y | ? |
| $H_2O_2$ in water ice | Y | Y | Y |
| Hydroxylated and hydrous minerals | Y | Y | Y |
| Halides | Y | ? | ? |
| Carbonates | Y | Y | ? |
| SH, $H_2S_n$ in water ice | | ? | X |
| $H_2SO_4 \cdot nH_2O$ in water ice | Y? | ? | ? |
| Sulfates (hydrated) | Y(?) | Y | ? |
| Organics CH, X-CN | | Y | Y |

Y = present. ? uncertain or suspected.

## Europa - overview

Europa is a differentiated moon comprised of silicate rock and possibly a small iron-nickel core. The ice:rock ratio of less than 10:90 is rather small by comparison to Ganymede or Callisto. The outermost shell over the silicate mantle largely consists of up to 5-6 km of water ice. Europa's surface temperature ranges from 86-132K, (Spencer et al. 1999). The surface is smooth and displays lineaments that are believed to be surface cracks from which subsurface brines entered onto the icy surface. Europa's surface materials include an abundant non-ice component at many locations which was suggested to be heavily hydrated minerals (McCord et al. 1998). There are hydrated silicates, salts, water ice, and $CO_2$, $SO_2$. The presence of water ice containing sulfuric acid hydrate, $H_2SO_4 \cdot nH_2O$ and/or hydrated salts, especially sulfates, is debated. However, the detection of NaCl salt is firm.

Spectra from the Galileo near infrared mapping spectrometer (NIMS) instrument indicated the presence of hydrated salts (McCord et al., 1998, 1999, 2001) in optically dark areas such as along the lineaments. These salts are expected when brines evaporate, thus whatever the thickness of the icy shell, there must be access of fluids from the ocean below to the surface. There seems to be consensus about the presence of a subsurface salty ocean (100km deep and with an estimated volume of 2-3 times the amount of water on Earth) that is kept liquid by heating through tidal flexing, however, two models vary about the thickness of Europa's icy surface layer. One set of models favors a 10-30 km thick surface layer of water ice, another posits an ice layer of only a few km. A thinner ice layer would enable brine fluids to break through more easily which would leave salt evaporite deposits behind. The discovery of localized NaCl salt deposits on the surface confirms the access of subsurface water to the surface (Trumbo et al. 2019). Further support for liquid water making it through to the surface are Hubble Space Telescope observations of water vapor plumes at Europa's south pole (Roth et al., 2014, Sparks et al., 2016, Jia et al. 2018).

The Galileo magnetometer results indicated the presence of a magnetic field for which induction within a brine ocean remains the best explanation as the field generating mechanism, a similar situation as for Ganymede. The relatively low frequency of impact craters suggests that there are still ongoing resurfacing processes. Europa's surface is covered with striations and cracks (lineaments) that are consistent with large-scale movement of ice.

## Ganymede - overview

The largest moon in the solar system (with a larger mass than Mercury or Pluto) is a differentiated object with a metallic core and silicate mantle which is covered by a massive 800-1000 km thick layer of mainly water ice. Models suggest about a 150 km thick icy surface under which a layer of liquid water exists, which in turn sits on layers of several high-pressure ice modifications. The subsurface saltwater ocean is estimated to be equivalent to at least that of the Earth. The top crust is covered by about 35% dark terrain and has 40% ice coverage including the polar ice caps.

Surface temperatures range from 85-150K (Hanel et al. 1979, Orton et al. 1996), and its surface is covered with water ice, hydrated salts, and organic compounds. McCord et al. (1998) found that the non-ice materials on Callisto and Ganymede must be much less hydrated and probably only hydroxylated than on Europa. Tosi et al. (2023) detected salts and organics on Ganymede's surface in spectra obtained with the JIRAM spectrometer onboard the Juno spacecraft. The identified salts include hydrated sodium chloride, ammonium chloride, sodium, and ammonium bicarbonates, and among the organics, possibly aliphatic aldehydes.

The presence of ammonium salts is of interest as the formation of these salts require a source of endogenic nitrogen, originally accreted to Ganymede as either $NH_3 \cdot H_2O$, $NH_4HCO_3$, $NH_4CO_2NH_2$ or as $N_2$ ice (or its clathrate hydrate) that was subsequently converted to $NH_3$ by some process (Prinn and Fegley 1981). With respect to sulfur, Tosi et al. (2023) also checked for ammonium sulfate, called mascagnite, $(NH_4)_2SO_4$. However, Tosi et al (2023) rule out the presence of Ca, Na, and $NH_4$ sulfates because their spectral shapes between 2.0 and 2.7 μm and beyond 3.5 μm are incompatible with the observed spectra. However, they found that the presence of bloedite, $Na_2Mg(SO_4)_2 \cdot 4H_2O$ is consistent with the observations. This sulfate is among the sulfates epsomite and bloedite which occur in situ or in water extracts of carbonaceous meteorites of type CI and CM, where sulfates formed as secondary terrestrial oxidation products (see below).

The JIRAM data analyzed by Tosi et al. (2023) did not reveal the presence of $SO_2$ (4.07 and 4.37 μm) nor the presence of exogenously produced hydrated sulfuric acid. However, in the darkened equatorial regions, Ligier et al. 2019 identified sulfuric acid hydrate and "chlorinated" salts. Organic compounds are also suspected in those regions. The salts and organics are susceptible to decomposition by radiolysis but they survive if they are shielded from radiation by Ganymede's magnetic field. The magnetic dipole field is possibly generated by tidal energy dissipation by convection within its fully or partially molten metallic core or within the putative subsurface salt ocean; the exact origin of the magnetic field is not fully understood. If the field depends on the presence of a salt ocean and tides for frictional heating to generate electric current by convective motion of brines, magnetic fields should also be present at the other "salt ocean-moons" Europa, Enceladus, and possibly Callisto, especially if accompanied by signs of energy transport from the interior as manifested by cryovolcanism and geysers.

### Callisto - overview

Callisto's density indicates an about equal contribution of icy and rocky materials to its interior. Models suggest it is differentiated into a metal-free (oxidized) silicate core overlain by an icy shell up to about 250 km thick. The ice layer may be deep enough to stabilize the high-pressure water ice V modification which is denser than the salt ocean between the ice V and the layer of normal ice I on the very top. Callisto is experiencing much less tidal heating than Europa and has no apparent geologic activity as the other Galilean satellites. Its old, heavily cratered surface has limited signs of resurfacing. About 60-80% of the surface is covered by dark, non-icy material that is interpreted as hydroxylated, but not hydrated silicates and organics whereas the icy patches are mainly water ice and $CO_2$ ice (McCord et al., 1997, 1998).

Carbonic acid (Johnson et al., 2004) has been suggested among Callisto's surface ices. The evidence for $SO_2$ and/or sulfanes is complicated (see below). The surface temperatures vary from 80K on the dark side to 155K at the subsolar point (Hanel et al. 1979). There is a tenuous $CO_2$-rich atmosphere (Carlson 1999) around $7.5{\times}10^{-12}$ bar, and molecular oxygen released by sputtering is suspected (Liang et al. 2005). Whether sulfur-bearing gases are present is unknown; the detection of either $SO_2$ or $H_2S$ could help to clarify whether $SO_2$ ice and/or $H_2S$ ice reside on the surface.

## Europa, Ganymede, Callisto: Follow the Sulfur

The Galilean satellites Europa and Ganymede share several properties regarding their sulfur chemistry: They have subsurface saltwater oceans, cryovolcanic plume activity, their icy surfaces contain salts and trapped recondensed icy $SO_2$ frost. Callisto, the chemical minimalist among the Galilean satellites, lacks signs of active surface geology, but shares Europa's and Ganymede's radiation induced surface chemistry. Common to all is their exposure to the Jovian magnetospheric energetic particle irradiation rich in S and O. This is an unavoidable factor affecting the sulfur cycle of the moons in the Jovian system. The sulfur cycle shown in Figure 3 applies to all water-bearing Jovian moons with water ice on their surfaces.

This section on the Jovian trio Europa, Ganymede, and Callisto describes common chemistry on their surfaces, up to the point where radiolytic production of surface sulfuric acid and sulfuric acid hydrate, $H_2SO_4{\cdot}nH_2O$ (n = 1-4, 6.5, 8) occurs.

All Jovian moons are exposed (partially only for Ganymede due to an internal magnetic field) to the strong magnetospheric radiation from Jupiter that, via interactions with Io's torus, transfers sulfur and other volcanic plume constituents (O, halogens, alkalis) from Io to the outer Jovian moons. The corotation of plasma makes the magnetospheric flux stronger on the trailing sides of the moons which often explains why abundances of substances with radiolytic production pathways are higher on the trailing hemispheres than the leading sides of the moons.

The magnetospheric plasma near the orbital planes of Europa, Ganymede, and Callisto has an average ion composition of about 14% $H^+$, 36% $\sum(O^+, O^{+2}, O^{+3})$, 47% $\sum(S^+, S^{+2}, S^{+3})$, and 3% $Na^+$ (from Figure 9 in Kim et al., 2020). Chlorine and K are also expected (see Io section above) but were not quantified in the study by Kim et al. (2020). While the flux composition is roughly constant, the flux density drops with distance from Jupiter and thus the time integrated transfer rate of S per surface area on Callisto is smaller than that on Europa and Ganymede.

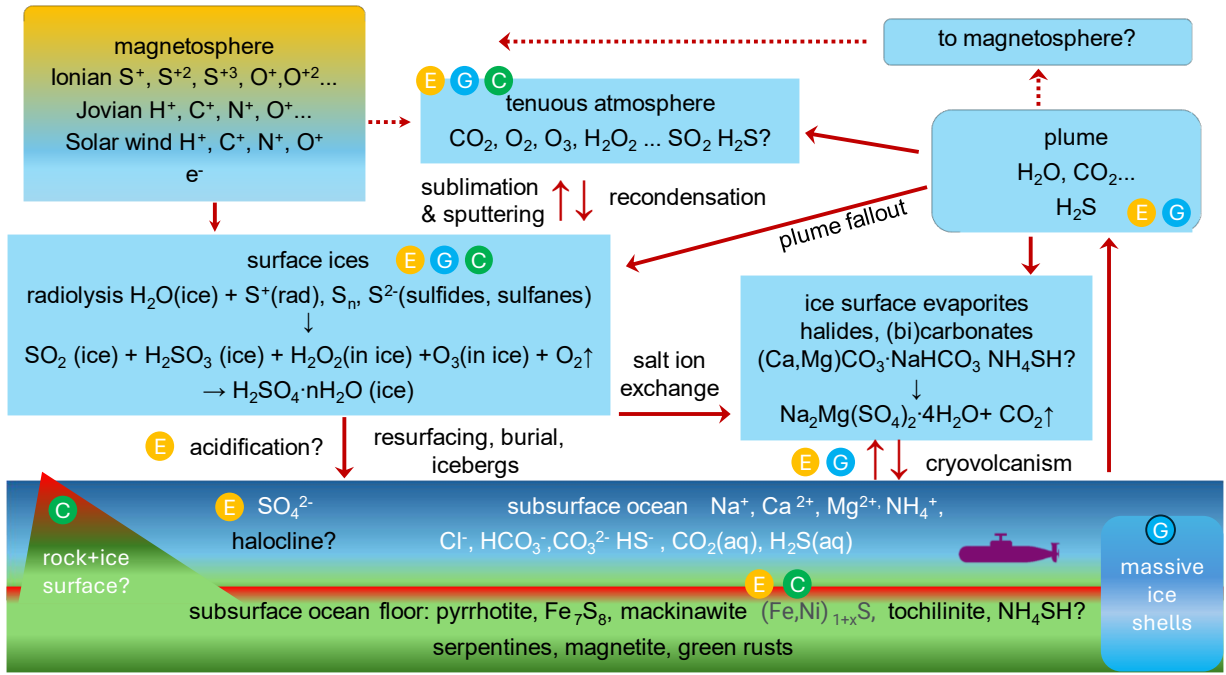

**Figure 3.** A schematic diagram of the sulfur cycle on the icy Galilean moons. Not all processes operate on all moons and the circles with E, G, and C indicate which processes seem to be more restricted to Europa, Ganymede, and Callisto, respectively. Ice radiolysis happens on all icy surfaces and reactions of radiolytically produced $H_2SO_4$ can transform brine salts to sulfates. Only for Europa and Callisto the subsurface ocean is likely to be in contact with the rocky ocean floor but not for Ganymede, where massive dense ice layers prevents ongoing exchange of the subsurface with the silicates. The ocean compositions seem to be all quite similar in the beginning, and Ganymede's ocean might be the oldest ur-ocean remaining since isolation by dense ice. Subsurface oceans on Europa and Callisto can evolve depending on silicate mantle outgassing into the oceans which may increase salinity over time. On Europa acidification from radiolytically produced $H_2SO_4$ may happen, which could lead to an ocean with compositional gradients and pH and development of a halocline. This diagram does not claim completeness and is an attempt to summarize observations, ideas and interpretations.

The presence of trapped $SO_2$ ice and $SO_2$ frosts are well established on Europa (e.g., Becker et al. 2020), Ganymede (McCord et al., 1997, 1998, Domingue et al. 1998.) On Callisto, $SO_2$ ice was seen at 4.07 and 4.37 µm (Hibbitts et al. 2000). The identification of an $SO_2$ absorption feature is problematic and it could be an artifact from ratioing spectra (Hendrix and Johnson 2008). Cartwright et al. (2020) found a stronger 4 µm band in spectra of Callisto's leading hemisphere and inferred the presence of disulfanide ($HS_2^-$ at 4.025 µm) and the presence of thermally altered sulfur and questioned the presence of $SO_2$ on Callisto. Ramachandran et al (2024) interpreted one of the features in their spectrum as $SO_2$. Overall, it seems that the presence of $SO_2$ ice in smaller quantities cannot be ruled out, but that the observational evidence for it remains moot.

The source of $SO_2$ on the satellite surfaces has been suggested as radiolysis of sulfur in water ice, or the decomposition of sulfate salts. One problem with the latter is that sulfates have to be generated somewhere at some point. If sulfate salts are left as evaporites from brines that came from the subsurface oceans, then from where did the oceans obtain the sulfates? The formation of sulfates requires an oxidizing agent, and rock-water interactions in absence of an oxidizer will not produce sulfates (more on that below).

Experimental studies of sulfur ion implantation into water ice yielded hydrated sulfuric acid, (Strazzulla et al. 2009, 2023) and energetic electron bombardment of elemental sulfur in water ice also produced the acid (Carlson et al. 2002). The production of sulfuric acid occurs via oxidation of sulfur within water ice (e.g., Carlson et al. 1999, 2002 and references therein, see also Loeffler and Hudson 2016). The net reactions can be written as

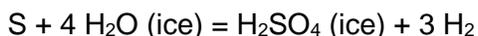

$$S + 4\ H_2O\ (ice) = H_2SO_4\ (ice) + 3\ H_2$$

and

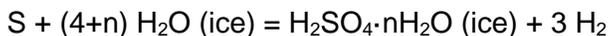

$$S + (4+n)\ H_2O\ (ice) = H_2SO_4{\cdot}nH_2O\ (ice) + 3\ H_2$$

The stepwise oxidation of elemental S proceeds via $SO_2$, $(SO_3)^{2-}$ and finally $(SO_4)^{2-}$. The source of the sulfur is thought to be either endogenic, i.e., from sulfur already present (S-H bonds are suspected in the water ice of Ganymede and Callisto, McCord et al., 1997, 1998, Cartwright et al., 2020), or exogenic, i.e., derived from Io's plumes and delivered as S ions in the magnetospheric plasma radiation. This radiation (<20 MeV electrons) also provides the energy to drive the sulfuric acid production in ice and enables oxidation of the implanted sulfur to $SO_2$ with O largely coming from $H_2O$, but O ions are also abundant in the plasma flux.

For abundant yields of $SO_2$ in the ice, a large and/or long flux of S is required. Radiolysis of $SO_2$ ice in the presence of water ice then becomes a source of sulfurous and sulfuric acid. The detected ions in the ices after irradiation in the lab are $H_3O^+$ and bisulfite, bisulfate, and sulfate anions $HSO_3-$, $HSO_4^-$, $SO_4^{2-}$ which upon warming convert to sulfuric acid hydrates $H_2SO_4{\cdot}nH_2O$ (n = 0-4) (Moore et al. 2007).

The critical step in the sulfuric acid production is the conversion of $SO_2$ to $SO_3$, better written as sulfite, $(SO_3)^{2-}$ or sulfurous acid $H_2SO_3$, which must be aided by reactive oxygen (denoted O*) such as in O, $O_2$, $O_3$, or $H_2O_2$ to overcome the high activation energy for the endothermic reaction $SO_2 + O* = SO_3$ (for the same reason industrial sulfuric acid production requires reactive oxygen and a catalyst for this step). This reaction requires about half of the energy that the subsequent exothermic hydration reaction $SO_3 + H_2O = H_2SO_4$ can supply.

The source of the reactive oxygen is provided by radiolysis of water ice and possibly also by radiolysis of $SO_2$ ice (see Io above). Energetic particle irradiation breaks up water ice and produces hydrogen peroxide via the net reaction $2H_2O\ (ice) \rightarrow H_2O_2 +H_2$. Radiolysis of water ice is indeed the most plausible explanation for making $H_2O_2$ ice (e.g., Gomis et al. 2004). Hydrogen peroxide is observed on Europa (0.24% maximum molar ratio of $H_2O_2/H_2O$, Hendrix et al. 1998, Carlson et al. 1999a, Hand and Brown 2013), on Callisto (0.15% maximum molar ratio of $H_2O_2/H_2O$, Hendrix et al. 1999) and in polar regions of Ganymede (Trumbo et al. 2023). Hand and Brown (2013) found lower $H_2O_2$ concentrations on the trailing hemispheres on Europa from Keck II observations which is somewhat surprising because that hemisphere is receiving most of the radiation. However, the non-icy components are less abundant on that hemisphere, which could affect $H_2O_2$ yields although this might not be the sole explanation. The radiolysis of ice in equatorial regions of Ganymede is limited by the induced magnetic field.

Another radiolysis product is ozone which was found on Ganymede (Noll et al. 1996), Callisto (Ramachandran et al 2024), and on the Saturnian moons Rhea and Dione (Noll et al. 1997). However, it has not been reported (yet) on Europa. Molecular oxygen $O_2$ is another water radiolysis product, and a decomposition product of $H_2O_2$ and $O_3$, thus $O_2$ can be expected to occur together with ozone and peroxide (e.g., Hall et al. 1995, Hendrix et al. 1999, see also Johnson and Jesser 1997 for photolytic production of $O_3$ from $O_2$ on Ganymede). The presence of $O_2$ in ice is consistent with absorption spectra on the surface of Ganymede (Spencer et al. 1995, Calvin et al. 1996, Migliorini et al. 2022), and Spencer and Calvin (2002) reported $O_2$ within ice for Europa (with larger $O_2$ abundances on the trailing hemisphere) and Callisto. The trapped $O_2$ is likely to be released by sputtering of water ice and contribute to some of the moons' tenuous atmospheres containing atomic O and $O_2$. As $O_2$ and $O_3$ occur together, ozone could also contribute to tenuous atmospheres.

It is instructive to compare the surface temperatures of Europa (86-132K, Spencer et al. 1999), Ganymede (85-150K, Hanel et al. 1979, Orton et al. 1996), and Callisto (80-155K, Hanel et al. 1979) to the melting points of $O_2$ (54.8K), ozone (80.7K), $H_2O_2$ (272K), $H_2S$ (191K), and $SO_2$ (197K). Taking these at face value, there is a possibility of trapped liquid $O_2$ and liquid ozone within the ices of Europa, Ganymede, and Callisto. In contrast, the detections of $O_2$ and $O_3$ ices imply temperatures below 55K. Of interest here is the possibility of liquid film percolation that could concentrate such oxidizing fluids and aid oxidation of S and $H_2S$ to $SO_2$ and sulfurous and sulfuric acids. At 60K the vapor pressure of $O_2$ is about 7 mbar and of $O_3$ only about 0.4 microbar, thus molecular $O_2$ is a more favorable candidate for atmospheric gas that can be transported and recondense. Molecular $O_2$ was found in the tenuous atmosphere (or exosphere) of Europa and Ganymede (Hall et al. 1995,1996). If $O_2$ is photolyzed in the atmosphere to $O_3$ (via the Chapman cycle, Johnson and Jesser 1997), upon condensation the more refractory $O_3$ could possibly act as an oxidizer on surface ices (including organics) in regions where radiolysis is not as effective. The possible existence of liquid $O_2$ and $O_3$ have been considered to explain some of the mysteries associated with distribution and occurrence of $O_2$ ice and $O_3$ ice on Europa(?), Ganymede and Callisto. Especially for Callisto, $O_2$ abundances in the atmosphere seem to be much larger than anticipated from radiolysis for which a full explanation still awaits (Carberry Mogan et al. 2023). If oxidation of sulfur species to sulfuric acid occurs, exothermic hydration of $H_2SO_4$ to $H_2SO_4 \cdot nH_2O$ (n = 1-4, 6.5, 8, 12) could become of interest. One could speculate that this further weakens the radiation damaged ice and would facilitate escape of trapped $O_2$ and $O_3$.

On Europa the observed distorted 1.5- and 2-micron absorption features of water ice in spectra from the NIMS (Near Infrared Mapping Spectrograph) on Galileo have been interpreted to indicate the presence of bound crystal water in hydrated salts, including sulfates, sulfuric acid, or both. The presence of hydrated sulfuric acid as a major surface component on Europa was inferred by Carlson et al. (1999b) who compared Galileo spectra with laboratory spectra of irradiated ices. According to their model, sulfuric acid is about 50 times more abundant than sulfur dioxide and sulfur allotropes. Carlson et al. (1999b) also interpreted the spatial correlation of visually dark surface material with sulfuric acid concentration as radiolytically altered sulfur polymers instead of evaporites. Apparently, hydrated sulfuric acid dominates Europa's trailing hemisphere, where the stronger irradiation naturally would increase sulfuric acid production whereas evaporite signatures are more concentrated on the leading hemisphere. Mishra et al. (2021), also using the NIMS spectra, concluded that water ice and sulfuric-acid-octahydrate are abundant but did not find any evidence for $SO_2$ and $CO_2$ ices in these spectra.

However, these same spectral signatures were also interpreted as originating from hydrated salts such as $MgSO_4 \cdot 7H_2O$ and $Na_2CO_3 \cdot 10H_2O$ (Fanale et al. 1999, 2000, 2001; McCord et al. 1998, 1999, 2001). These salts appear in dark regions near lineaments and are plausibly explained as evaporites from brines that broke through the ice from a subsurface ocean.

The truth might be in-between. For Europa, Orlando et al. (2005) found the best agreement of NIMS Europa spectra with multi-component mixtures of sodium and magnesium sulfate salts plus sulfuric acid in approximate molar ratios. 50:40:10 of $MgSO_4$:$Na_2SO_4$:$H_2SO_4$. Whether these are to be interpreted as pure endmembers, or solid-solutions of cation-bearing salts coexisting with "pure" endmember $H_2SO_4$; or as bisulfate ($HSO_4^-$) bearing salt mixtures remains unclear.

The Galileo NIMS spectra have too low spectral resolution to distinguish between the hydrated salt and hydrated sulfuric acid interpretations. Brown and Hand (2013) obtained high resolution spatially resolved spectra of Europa between 1.4 and 2.4 µm using adaptive optics and the OSIRIS integral field spectrograph at the W. M. Keck Observatory. They found a new spectral absorption feature at 2.07 µm in the equatorial region of the trailing hemisphere but not in its polar regions, nor was it found on the leading hemisphere. The absorption feature is strongly correlated with the presence of $SO_2$. The trailing hemisphere receives stronger magnetospheric radiation and the spatial association of sulfate and $SO_2$ points toward radiolysis as a major driver of the surface sulfur chemistry.

Brown and Hand (2013) took lab spectra of a large number of possible ices, salts, and brines in different forms, and only identified epsomite as a matching compound for the 2.07 µm absorption on the trailing hemisphere. Using the spectral libraries for ices, salts, and brines from Dalton and Pittman (2012) they were able to fit the overall spectrum between 1.4 and 1.8 µm and 2 – 2.4 µm with a mixture of about equal amounts of hydrated sulfuric acid and magnesium sulfate brine, but an additional, (not dark) neutral material was also needed, especially for obtaining a good continuum match of the observed spectrum at 1.4 and 1.8 µm.

Trouble for the sulfate story came from linear spectral modelling of the other, evaporite-rich leading hemisphere. Brown and Hand (2013) obtained a best least-squares fit to the observations between 1.4 and 1.8 µm and 2 – 2 4 µm with a mixture of 35.4% sulfuric acid hydrate, 18.2% hexahydrite, 17.9% mirabilite, 7.7% water ice of 100 µm grain size, and 20.6% water ice of 250 µm grain size. However, hexahydrite should also produce a small absorption at 1.6 µm, and mirabilite one at 2.18 µm which were not observed. Thus, while the spectral shape points towards hydrated salts their identity as sulfates remains unclear on the leading hemisphere. Brown and Hand (2013) therefore considered the possibility that the evaporites consists of other salts such as halides.

Still, the sulfate salts on the trailing side have to come from somewhere, and if they are not in the same brines that fed the leading hemisphere, they must made at location. Radiolysis only leads to hydrated sulfuric acid, but sulfates can be made from it if other suitable brine salts are present. Salts of weaker acids such as carbonates, bicarbonates are ideal candidates for replacement reactions if in contact with sulfuric acid:

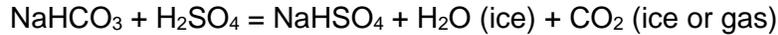

$$NaHCO_3 + H_2SO_4 = NaHSO_4 + H_2O \text{ (ice)} + CO_2 \text{ (ice or gas)}$$

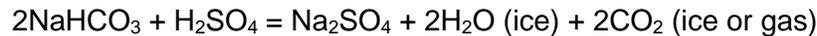

$$2NaHCO_3 + H_2SO_4 = Na_2SO_4 + 2H_2O \text{ (ice)} + 2CO_2 \text{ (ice or gas)}$$

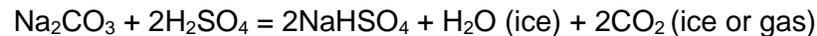

$$Na_2CO_3 + 2H_2SO_4 = 2NaHSO_4 + H_2O \text{ (ice)} + 2CO_2 \text{ (ice or gas)}$$

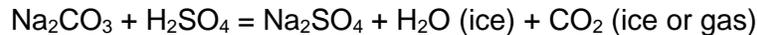

$$Na_2CO_3 + H_2SO_4 = Na_2SO_4 + H_2O \text{ (ice)} + CO_2 \text{ (ice or gas)}$$

Similar reactions could occur with ammonium bicarbonate if $NH4^+$ is not destroyed during radiolysis.

The production of $CO_2$ may contribute to the tenuous atmospheres on Jupiter's moons. In this context, destruction of any organic carbonaceous material by $H_2SO_4$ is another source of $CO_2$ that could also be important on Callisto, which has the most dense and $CO_2$-rich atmosphere among the icy moons (Carlson 1999, Cartwright et al. 2023).

The cryovolcanic plumes and observed salts on the surfaces are generally taken as evidence that there must be a salty subsurface ocean from which the salts are sourced. The regional occurrence of the observed sulfates in particular does not rule out the possibility that sulfates form from other salts on the surface by reaction with produced sulfuric acid. Of course, then there still has to be a source for the other salts as a cation source. Let us summarize what salts other than sulfates are actually observed on the icy surfaces, and what can be said about the presence of any brine salts that would point to the composition of any subsurface water bodies. Within a salty subsurface ocean (see below) the candidate salt formers are chlorides, sulfates and sulfides, and carbonates, this just follows from considerations of elemental abundances and geochemical principles (see also extensive works by Zolotov on this subject). Brine cations are alkalis (mainly Na), alkaline earths (Ca, Mg), and ammonium $NH4^+$.

For Europa, only the detection of halite (NaCl) seems to be firm, and spectral signatures of carbonates are not reported as of the time of writing. Hydrated salts, identified by distorted crystal water can include chlorides, (bi)carbonates, and sulfates, or even double salts, but the data are inconclusive about the cation and anion species of the hydrated salt.

For Ganymede, there is evidence for ammonium salts but not ammonium sulfate (Tosi et al. 2023, see above). The spectral evidence points to ammonium as chloride and bicarbonates. The other inferred salts are hydrated NaCl and $NaHCO_3$. All elements for these salts are provided to the surface independently, and these salts could form in situ on the surface just like sulfuric acid would. The interesting aspect is that they may not need a salty ocean as a source. Like sulfur, Na and Cl are also supplied from Io via the magnetospheric flux, and there is also some N (courtesy of Jupiter). As described above, another source of N could be ammonia hydrates or perhaps $NH_4SH$ already among the surface ice as is $CO_2$ on Ganymede. Whether ammonium and/or Na chlorides and carbonates actually form after ion implantation is not clear and needs to be evaluated, but it also cannot be ruled out. Note that radiolysis is acting on the top cm of material on the surface ice, but so is spectroscopy and an argument for massive amounts of salt (and or the flux being too small) may not hold.

Proof that brine salts (from a subsurface ocean) are present must come from other cations such as Mg and Ca. For Ganymede, Tosi et al. (2023) also found evidence for $Na_2Mg(SO4)_2 \cdot 4H_2O$ (bloedite) as that gave the best match for the spectral shape. Making this salt requires a source of Mg, for which the magnetospheric plasma may not be a sufficient source. If indeed present, this mineral must have come from a brine residue, or was produced by reaction of some other

Mg salt by reaction with $H_2SO_4$ of radiolytic origin. Such an origin was suggested for $MgSO_4$ from $MgCl_2$ and $H_2SO_4$ on Europa by Brown and Hand (2013).

For Callisto the possibility of a salty subsurface ocean is frequently discussed, but evidence for actual salts on the surface are lacking. A recent DPS abstract reports on the stronger $CO_2$ ice absorptions on Callisto's trailing hemisphere but does not mention carbonates.

## Saturnian Satellites

There are 146 satellites orbiting Saturn. The seven major moons, in order of distance from Saturn, are Mimas, Enceladus, Tethys, Dione, Rhea, Titan, Hyperion, and Iapetus. They belong to the 24 regular satellites with prograde orbits and (except Iapetus) have minimal inclinations to Saturn's orbital plane. With respect to sulfur, there are detections of $H_2S$ but not sulfates in Enceladus, and for Titan little is known as described below. We refer the reader to the books edited by Müller-Wodarg et al (2014) and Schenk et al. (2018) for a more general overview of Saturn's major satellites.

### Enceladus

Enceladus is the smallest planetary object within the solar system with still ongoing endogenic geologic surface activity. Enceladus has several smooth terrains covered with tectonic fractures, troughs and lineaments, contrasting heavily cratered (thus older) regions. The thick icy surface has the highest reflectivity in the solar system. The abundant surface ices are water ice followed by $CO_2$ trapped in ice (Parkinson et al., 2007).

The south polar region has endogenic activity. It contains several sub-parallel creased features called sulci (furrows and ridges; or colloquially "tiger stripes"). In this region, jets of water vapor and dust cumulate into a plume. The spectra collected during the CASSINI mission with the Composite Infrared Spectrometer (CIRS) showed enhanced surface temperatures in this region (Spencer et al. 2006, Porco et al. 2006). The increased temperatures at the pole are likely caused by tidal heating resulting from a 2:1 orbital resonance with Saturn's moon Dione. Current observations and models suggests that Enceladus has a larger sea under its southern pole, and possibly a global sub-surface ocean (e.g., Iess et al. 2014) where hydrothermal activity is ongoing.

The ejecta feeding the plume originate from the vents at the "tiger stripes" but ejecta fall-back and resurfacing occurs which recycles material to the cryovolcanic source reservoir. Brown et al. (2006) found that the ice in the area of the "tiger stripes" is crystalline indicating younger deposits than the amorphous ice seen in other regions. In conjunction with plume observations, Villanueva et al. (2023) also found high ice crystallinity on the trailing hemisphere. Resurfacing also adds new materials to the source reservoir such as organics from infalling meteoritic and cometary matter which gets buried along with the recycled ice. This is important as it provides a replenishment of carbon in $CO_2$ and organics lost into Enceladus' plume and torus (see below).

One suspected icy component at the south polar region is $H_2O_2$ identified by a 3.5-micron feature on top of a water absorption plateau in spectra taken by CASSINI's Visual and Infrared Mapping Spectrometer (VIMS) (Newman et al. 2007) However, Loeffler and Baragiola (2009) found that this assignment is probably incorrect and could be due to water ice. Hodyss et al. 2009 suggest that the 3.5-micron feature is methanol instead. Villanueva et al. (2023) also observed a subtle 'plateau' at 3.5 μm in spectra from the James Webb Space Telescope near-infrared spectrograph (JWST–NIRSpec) and find that it could be consistent with radiolytic production of $H_2O_2$ on $H_2O$ ice but that the spectral identification remains uncertain. As above for the Jovian satellites, the finding of $H_2O_2$ in the south polar region would be important because $H_2O_2$ serves as an oxidizing agent. Some oxidizer is required if organics are converted to $CO_2$ or carbonates with the plume source reservoir by subsurface hydrothermal activity. Villanueva et al. (2023) searched the south pole region for but did not detect ices of $NH_3$ and its hydrates which are also not seen in other regions.

Cryovolcanism continuously expels matter from Enceladus at an estimated rate of 300 kg/s into large plumes extending far beyond Enceladus - recent JWST observations reveal a plume close to 10,000 km in size equal to about 40 times the radius of Enceladus (Villanueva et al. 2023). The ice and salts expelled by the plumes form a largely neutral torus rich in oxygen and hydrogen along the moon's orbit. Matter in the torus is a major source for Saturn's E-rings (Postberg et al. 2009, 2011), and also for ions in Saturn's magnetosphere. The HERSCHEL observation of the water torus (Hartogh et al. 2011) and JWST results (Villanueva et al. 2023) also suggest that the torus can provide water to Saturn's ring systems and that fluxes are even sufficient to supply $H_2O$ to Titan and Saturn's upper atmosphere in observed quantities there.

The plume from the subsurface ocean contains a variety of gases. Villanueva et al. (2023) made detailed observations of the plume and torus with JWST. They detected $H_2O$ in the plume but did *not* detect molecular emissions of $CO_2$, CO, $CH_4$, $C_2H_6$ and $CH_3OH$ across the plume. However, the Cassini Ion and Neutral Mass Spectrometer (INMS) detected (in decreasing abundance) $H_2O$, $NH_3$, CO, $CO_2$, $CH_4$, HCN, $C_2H_2$, $C_3H_6$, and $H_2$ plus an assortment of other organics and N-bearing gases (Peter et al. 2023; some of the CO could be an instrumental artifact and impact generated from $CO_2$). Their $H_2S$ detection as an upper limit at mass 34 is tentative and could also stem from $PH_3$ or $H_2O_2$. Sulfate salts have been looked for, but none have been identified. In light of the subsequent Na-phosphate discovery in the plume (Postberg et al. 2023, and chemical considerations, the presence of $PH_3$ as the mass 34 source is less likely. Sulfates were looked for but seem to be absent and are mutually exclusive with $H_2S$, which is the plausible source for the mass 34 signal.

The evidence for Na-phosphates in the plume is expected from aqueous alteration in alkaline and $CO_2$-rich salt waters which modelling suggests for Enceladus' Ocean (Fifer et al. 2022, Glein et al. 2015). The same applies to heavily aqueously altered carbonaceous CI-chondrites which are a natural laboratory and show the carbonates, phosphates, halides and other salts from such fluids, and for which a detailed assessments of aqueous alteration were done (see references in the ocean section). All these thermochemical models show that sulfates are not expected to form unless a strong oxidizing agent is present. If there is an oxidizing source with Enceladus' ocean, anhydrite and gypsum would be the only expected sulfates. The solubility of $CaSO_4$ is low, and most of it would sit on the ocean floor, thus plume concentrations of sulfate, if any, would be extremely low. Zolotov (2007) presents detailed models for the evolution of Enceladus' chemistry.

## Titan

Titan is Saturn's largest moon, and the second largest moon in the Solar System. It contains most mass (96%) of all Saturnian moons combined. It has an $N_2$-rich atmosphere also containing $CH_4$ and other hydrocarbons, and hydrocarbon lakes. Nixon et al. (2013) reported an upper limit for $H_2S$ of 330 ppb by volume. The reduced chemistry suggests that $H_2S$ should be stable but the vapor pressure of $H_2S$ ice is quite low (Giauque and Blue 1936) and if present, most $H_2S$ is easily condensed out.

The discovery of a 5-micron bright spot (Barnes et al. 2005) stirred interest in models for the possible compounds causing it. The Cassini/Huygens mission revealed an about 500 km wide bright spot within the highly reflective region of "Xanadu" on Titan. This spot appears bright in all wavelengths but particularly bright at the 5 micron ("red") spectral window (Barnes et al. 2005). The nature of this bright material remains unidentified. It is not associated with any temperature variations that could point to cryovolcanism.

Fortes et al. (2007) suggested ammonium sulfate (or its tetrahydrate) as a candidate for the 5 μm-bright spot, or for the unidentified IR-blue component at the Huygens landing site. They make a case that the infrared reflectance spectrum of ammonium sulfate is a plausible candidate for the 5 μm-bright material on Titan's surface. They propose the reaction

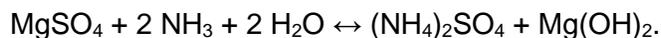

$MgSO_4 + 2 NH_3 + 2 H_2O \leftrightarrow (NH_4)_2SO_4 + Mg(OH)_2$.

For this reaction Mg sulfate has to come from somewhere and similar problems as described above arise. They reference high pressure and temperature hydration experiments of chondritic rocks (Scott et al. 2002) and claim that work "confirms that sulfur is expressed as sulfate at pressures up to 1.5 GPa and that higher pressures and temperatures favor the formation of sulfides". However, Scott et al. (2002) do not report formation of sulfate and found pyrrhotite over a wide range of oxygen fugacities from nickel – nickel oxide (NNO) to magnetite – hematite (MH), which is quite oxidizing, under their experimental P, T conditions.

Sulfate production requires an oxidizing step if produced from originally accreted $H_2S$, sulfides or elemental sulfur. In their model Fortes et al. (2007) postulate a chondritic source (which always contains Fe sulfide), plus water, and $CH_4$ and $NH_3$ from ices. This does not include an oxidizer strong enough for sulfide to sulfate conversion. If an oxidizing agent could be found, there is also the problem that hydrocarbons would oxidize before sulfur compounds.

The presence of $N_2$, $NH_3$, and abundant polymeric organic matter, hydrocarbons, and $CH_4$ rules out oxidized chemistry for sulfur. Sulfur speciation in a $H_2O$-$NH_3$ fluid is more likely to be in the form of sulfides such as $S^{2-}$ or $HS^-$ and as $NH_4SH$ or $(NH_4)_2S$ than sulfates. At the extremely low temperatures at the surface-atmosphere interface on Titan, water ice is essentially inert to reactions with the more volatile organics. We can only speculate about the presence of $CS_2$ dissolved in the $CH_4$ lakes or among the organic precipitates on Titan's surface. From the overall chemical point of view, thiazyl compounds might be attractive alternatives in the N-and C-rich environment.

## Uranian and Neptunian Satellites

There are no reports yet on any observed sulfur bearing compounds for the Uranian and Neptunian moons.

## Sulfur below the ice caps: Subsurface oceans and rocky portions

We now come to the question about the composition and evolution of the putative subsurface oceans, and much of what follows also applies to any subsurface oceans of moons in the Saturnian system (see below). The chemistry of salty liquid water under the surface ice has been studied extensively using geochemical principles and cosmochemical considerations (see

Kargel 1991, Kargel et al. 2000, Fanale et al. 2001, Glein et al. 2008, 2015, Glein and Shock 2010, Zolotov 2007, 2012, Zolotov and Shock 2001, 2003, 2004, Becker et al. 2024).

We cannot repeat all the details here and only describe some aspects that we see pertinent to the sulfur chemistry, there are also many new developments and reports as this is a highly active area of research and exploration.

The key questions are what can dissolve into an ur-ocean from the initially accreted rocks and ices, and how will this fluid evolve. This depends on the initial and final ice to rock ratios, which are even today obviously vastly different for Europa, Ganymede, and Callisto. Although water ice is the major accreting ice, depending on P, T conditions in a circumplanetary disk, there is a good change of co-accretion of $CO_2$, $CH_4$, $N_2$ and/or $NH_3$, but also $H_2S$ bearing ices, as noted earlier. How much of the volatile ices can be retained depends on how deep and well they are mixed with rocky materials as their volatilization during water ice melting is unavoidable. A frozen ice cap on the top would help to make a close hydrothermal system, keep gases inside, and pressure build up (gas-pockets?) would also be supportive for dissolving gases such $CO_2$, $NH_3$, and $H_2S$ in liquid water. These gases also dissociate which is important for setting the initial pH of aqueous solution and attacking the rocky portions in proto moons to get the cations into the water. On differentiated moons, temperatures must have been higher which means that we can expect internal(!) steam atmospheres within these objects during and after accretion which is very effective of fractionating rocky elements from silicate melts (if once present) or solid minerals into an aqueous fluid (e.g., Fegley et al. 2016 and references therein). In a similar manner as $CO_2$, $NH_3$, and $H_2S$ control the pH of an aqueous solution, these gases may control the redox state of the internal steam atmosphere. Such atmosphere would certainly attempt to outgas, and if so, a large portion of (gas and liquid) fluids may end up as salty, liquid water, depending on outgassing and cooling rates, which are all pretty model dependent. With a strong temperature gradient from top to bottom within a moon, water ice freeze-out at the top occurs naturally, enriching any remaining water in dissolved salts. If interior heating persists (aided by tidal heating), such internal steam atmospheres would be sustained, giving rise to interesting scenarios where continuous extraction of salty components end up in the aqueous layer.

The high temperatures are required to make metallic cores and silicate portions on Europa, and possibly Ganymede (see Soderlund et al. 2020 for interior models), as the minimum temperature of the Fe-FeS eutectic is around 1200K and melts are needed for the density separation of metal from silicate (partial) melts. This also shows that there is the need to prevent the oxidation of accreted metal by $H_2O$ (or to have a reducing agent in the case that the iron in the accreted rocky portion was already oxidized

This is one constraint for either the initial ice:rock ratio, for the requirement of heterogeneous accretion, or for the stratigraphy of reaction zones during differentiation (mixing timescales). If ice and rock accreted together, rapid separation of volatilzed ice into upper layers might prevent oxidation of metal in lower layers and a core could form. Core formation is important for the sulfur chemistry because removal of FeS together with metal decreases the amount of sulfur in the silicates and thus reduces sulfur availability for any aqueous phase.

## Making a salty ocean

Traditionally many planetary scientists have looked at the cosmochemical evidence from meteorites as a starting point for results of aqueous alteration processes in the early solar system, and when applied carefully, this approach is still valid. Evolutionary models for the moons commonly start with accretion of dry chondritic material, which is often just called "rock". consisting of water-free silicates, sulfide (FeS) and FeNi alloys) and ices (mainly water but also $N_2$, $NH_3$, $CO$ and $CH_4$ ices, their hydrates, and clathrates, and organics (e.g., Kargel, 1991; Kargel et al. 2000, Fanale et al., 2001; Kargel et al., 2000, Zolotov 2007, 2012, Zolotov and Kargel 2009, Zolotov and Shock 2001, 2003, 2004).

The proportions of "rock" and "ice" are free parameters and often chosen to approximate the densities of the moons; sometimes significant contribution of organics are also considered. The density of ice is approximated by that of water ice (0.98 g/cm$^3$) and average ordinary and enstatite chondrites have densities of about 3.6-3.8 g/cm$^3$ depending on total metal to silicate ratios; carbonaceous chondrites with low water contents (<3 mass%) have similar densities. Carbonaceous chondrites with larger water contents (bound to hydrous silicates) and no or little metal (after conversion to magnetite) have densities of 2.23 g/cm$^3$ (CI-chondrites) and 2.71 g/cm$^3$ (CM-chondrites). The condensable elements in solar composition material give an ice:rock mass ratio = 1.1 (Table 11 in Lodders 2003). This ratio is strongly influenced by the adopted solar oxygen abundance as this determines how much water ice can form, and older solar compositions with higher oxygen abundances allowed ice:rock ratios of up to two. However, the ice:rock ratio in the early solar system was also a function of heliocentric distance and the bulk ratio from solar composition is just a plausible starting point for models.

The ice to rock ratio that is indirectly reflected in the moons' densities is only a lower limit of that what originally accumulated into the moons. Aqueous oxidation of metal and silicate hydroxylation and hydration reactions will change the balance of metal available for a core, and "free" water for making oceans and icy shells. The low densities Callisto and Enceladus (Figure 2) and the inferred absence of a metallic cores in them suggest that sufficient ice must have accreted to oxidize preexisting metal whereas the low density of Ganymede with a putative metallic core is dominated by the massive amounts of water ice which indicates that a quite high ice: rock ratio was present at its formation location.

Contemporary with the loss of some part of the "ice fraction" by the oxidation of iron to magnetite via the net reaction

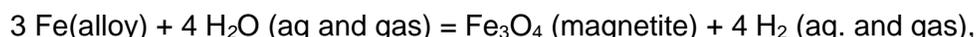

$$3\ Fe(alloy) + 4\ H_2O\ (aq\ and\ gas) = Fe_3O_4\ (magnetite) + 4\ H_2\ (aq.\ and\ gas),$$

another larger fraction of the water ice converts originally "dry" rocky Mg,Fe-chondrite silicates or silicates in the differentiated silicate mantles such as olivines and pyroxenes to phyllosilicates (see below). Iron and Mg are the major rock forming cations (aside from silicon), and the loss of water as separate liquid and icy phase on the moons would be considerable.

The aqueously altered carbonaceous CI- and CM chondrite are good analogs for mineral assemblages that are left after aqueous alteration and loss of water (but see also limitations regarding the sulfur chemistry in these meteorites below). Upon contact with water, halides in chondritic rock can partition into aqueous phase. Halides such as NaCl are the most likely carrier of halogens in primitive chondrites next to (metamorphic) phosphates (see Lodders & Fegley 2023 for halogens in chondrites). The presence of carbonates on the icy moons mimics that in CI and CM chondrites where carbonates precipitated from aqueous fluids within a few million years after solar system formation. It requires that $CO_2$ for carbonate formation was present in all cases. Detailed thermodynamic modelling of aqueous fluids on the moons indicates that early aqueous fluids are alkaline and contain Na$^+$, Cl$^-$, and HCO3$^-$ (Zolotov 2007, 2012, Glein et al. 2015, Fifer et al. 2022). The computations also showed that these fluids coexist with Mg-rich phyllosilicates (saponite, serpentine), calcite, magnetite, pyrrhotite, and Ni sulfide(s). The occurrence of these minerals is remarkably similar to that seen in CI and CM chondrites which is one of the many reasons why these meteorites are used as natural laboratories for low temperature aqueous planetary processes. There is also a sulfide phase in CI and CM chondrites that is not widely recognized but which could be of importance for the sulfur cycle of the moons.

In order to understand the sulfur chemistry during aqueous alteration of initially dry chondritic rock, we also need to know the reaction of the other major players which influence the sulfur distribution and speciation. Before doing so, we address the caveat of the incorrect assumptions that the entire mineralogy in carbonaceous chondrites seen today is still that of chondritic rock after aqueous alteration on the parent asteroid.

The sulfates in CI- and CM- chondrites, in particular the hygroscopic hydrated Mg-sulfates, can be secondary alteration products made in museums in the presence of moist air (Gounelle and Zolensky 2001, King et al. 2020). They can also be responsible for about 6-10% of terrestrial water in CI-chondrites and the current water:rock ratios of CI-chondrites is not necessarily characteristic of the ice:rock of its parent asteroid. The current mineralogy with sulfates in CI- and CM-chondrites is not necessarily representative of the original sulfur speciation of CI- and CM-chondrites. Instead of sulfates as initial products of aqueous alterations, the focus should be on sulfides in addition to halides and carbonates (see discussion on pp. 79-93 of Lodders and Fegley 2011). Experiments of extracting soluble salts from CI and CM-chondrites and equating the water compositions to the original fluid compositions coexisting with the hydrous silicates also have complications. For example, Fanale et al. (2001) found abundant sulfate extracted from the CM-chondrite Murchison in distilled boiling water. However, some part of these extracted sulfates could easily have formed during boiling as the water used was not degassed from air and no inert gas was lead through the solution or kept over it. While care was taken to remove any rusty spots (terrestrial oxidation products) from the CM-meteorite sample (Murchison fell in 1969), fine rust is hard to remove and can serve as sulfide oxidizer. It can be shown that fresh CM chondrite falls contain lower amounts of soluble sulfates. The sulfate content of CI-chondrites also increases with meteorite residence time on Earth, and the original sulfate content, was much lower – we estimated that only 1/5[th] of the total sulfur was present in sulfate form when the meteorite fell (see Figure 2.19 in Lodders and Fegley 2011). Thus, the currently observed sulfur speciation in CI and CM chondrites, and possibly that of other carbonaceous chondrites that experienced aqueous alteration on their parent bodies, is not representative for the original sulfur speciation resulting from aqueous processing that occurred in the absence of oxidizing gases. Therefore, early studies that adopt the sulfate abundances in CI or CM chondrites as *initial* proxy compositions in their models (e.g., Hogenboom et al. 1995, Kargel 1991, Kargel et al. 2000, Fanale et al. 2001) may require re-assessment about the sulfur chemistry (see also McKinnon and Zolensky 2003).

Isotopic data provide some insight into the origin of the sulfates in CI- and CM-chondrites. Airieau et al. (2005) studied the oxygen isotopic composition of water-soluble sulfates in CI and CM chondrites. The oxygen isotopes in the CM sulfates are consistent with origin on an asteroidal parent body but the Orgueil and Ivuna CI sulfate oxygen isotopes are indistinguishable from terrestrial. Airieau et al. (2005) caution their data do not rule out an asteroidal origin for the sulfates in CI chondrites.

However, we still can get a good lesson from the sulfates of carbonaceous chondrites: They show us what happens to a hydrated silicate and salt mineral assemblage once exposed to an oxidizing agent – quite similar to that what is happening to deposited evaporite salts on the moons' icy surfaces, or even surface rocks if exposed (Callisto?).

We now look at aqueous alteration products in addition to the magnetite already mentioned. A detailed description of the various alteration reactions for silicates to hydrous and hydroxylated silicates, to the green-rust silicates (containing carbonates), and to silicates containing sulfur is in chapter 2 of Lodders and Fegley (2011); for the formation of hydrated and hydroxylated silicates see also e.g., Sleep et al. 2004, Zolotov 2014, Holm et al. 2015, Sekine et al. 2015:

The **serpentinization** reaction of forsterite, the Mg-endmember of olivine, can be written as

$$2\ Mg_2SiO4\ (forsterite) + 3H_2O = Mg_3Si_2O_5(OH)_4\ (Mg\text{-chrysotile}) + Mg(OH)_2\ (brucite)$$

The analog reaction for fayalite, the Fe-rich endmember is

$$2\ Fe_2SiO4\ (fayalite) + 3H_2O = Mg_3Si_2O_5(OH)_4\ (Fe\text{-chrysotile}) + Fe(OH)_2$$

Ferrihydrite, an oxidation product of ferrous hydroxide with the formula $Fe_2O_3 \cdot 5H_2O$ is closely associated with magnetite and phyllosilicates in CI chondrites. The trivalent ferrihydrite forms from the reaction above but iron hydroxides with intermediate valences between $Fe^{2+}$ and $Fe^{3+}$ result from metal and sulfide oxidation.

The analogous reaction for enstatite, the Mg-endmember of pyroxene leads to talc,

$$5\ MgSiO_3\ (enstatite) + H_2O = Mg_3Si_2O_5(OH)_4 + Mg_2SiO_4\ (forsterite)$$

Serpentine $(Mg,Fe)_3Si_2O_5(OH)_4$ is the solid-solution of Mg-endmember chrysotile and Fe-chrysotile where Mg and Fe can exchange places in the crystal structure. This couples magnetite formation to serpentinization with release of $H_2$:

$(Mg_xFe_{1-x})_3Si_2O_5(OH)_4\ (chrysotile) + 3y\ Mg(OH)_2\ (brucite) = (Mg_{x+y}Fe_{1-x-y})_3Si_2O_5(OH)_4\ (chrysotile) + y\ Fe_3O_4\ (magnetite) + 2y\ H_2O + y\ H_2$

Reactions like this producing $H_2$ can lead to different outcomes in the subsequent alteration processes, depending on whether $H_2$ is lost from the system or not (see Zolotov and Mironenko 2007, Zolotov and Kargel 2009 for detailed modelling). The hydration of pyroxene and olivine is fast and exothermic, and makes the solution slightly alkaline, which favors $CO_2$ dissolution. If $H_2$ reduces $CO_2$ via

$$CO_2 + H_2 = CH_4 + H_2O,$$

the decreased partial pressure of $CO_2$ also reduces $(CO_3)^{2-}$ and $(HCO_3)^-$ concentrations in solution. Although this reaction may require a catalyst to reach equilibrium, the $CO_2/CH_4$ ratios such as observed e.g., in plume material of Enceladus, can still serve as an indicator for the $H_2/H_2O$ ratio and thus give limits to the redox state (see discussions in Zolotov 2007, Zolotov and Kargel 2009). Changes in the carbonate and bicarbonate anion concentrations also affect solution equilibria of cations of Na, Ca, Mg, and Fe and precipitation of carbonates such as calcite $(CaCO_3)$ and magnesite $(MgCO_3)$.

The net transformation of olivine to serpentine and carbonates which are observed in CI chondrites may also occur on the ocean-bearing satellites:

$$2\ (Mg,Fe)SiO_4 + 2H_2O\ (aq) + CO_2\ (aq)$$

$$= (Mg,Fe)_3Si_2O_5(OH)_4\ (chrysotile) + (Mg,Fe)CO_3\ (breunnerite)$$

Or written as a dissociation equilibrium

$$2\ (Mg,Fe)SiO_4\ (olivine) + 2H^+ + H_2O\ (aq) + HCO_3^- = (Mg,Fe)_3Si_2O_5(OH)_4\ (chrysotile) + (Mg,Fe)^{2+} + 2CO_3^-$$

In $CO_2$-poor fluids $FeCO_3$ (siderite) does not precipitate and iron may precipitate as hydroxide instead, which is an important way to separate Mg from Fe. However, divalent Fe may stay in solution if $NH_4^+$ is present, which is important because $NH_4^+$ salts were found in early studies of CI-chondrites, and there is evidence for ammonium salts on some of the moons' surfaces. The reaction

$$NH_4^+ + CO_3^{2-} = NH_3 + HCO_3^-$$

can transform insoluble carbonates to soluble bicarbonates. If $NH_3$ is driven out of the solution, it also has implications for the stability of $NH_4SH$.

The reactions when accreted troilite, FeS, is exposed to an aqueous fluid are not well studied. Troilite is stoichiometric FeS whereas pyrrhotite, $Fe_7S_8$ also written as $Fe_{0.875}S$, is the observed iron sulfide in CI chondrites. This sulfide occurs isolated as isolated hexagonal platelets that appear to have precipitated from solution.

Zolotov and Kargel (2009) describe anoxic oxidation of troilite and pyrrhotite through interactions with $H_2S$. The dissolution of $H_2S$ is favored in acidic solution in systems with $H_2$ removal. For these reactions to operate, an independent source of $H_2S$, possibly accreted with other ices, would be required:
$(1{-}x)FeS$ (troilite) $+ xH_2S$ (aq) $= Fe_{1-x}S$ (pyrrhotite) $+ xH_2$

And subsequently sulfidation to pyrite in $H_2S$ rich systems

$Fe_{1-x}S$ (pyrrhotite) $+ (1{-}2x)\ H_2S(aq) = (1{-}x)FeS_2$ (pyrite) $+ (1{-}2x)H_2$

Pyrite is not reported for CI and CM chondrites but remains a possibility for the moons. However, pyrrhotite is not the entire sulfur inventory in CI and CM chondrites. Even accounting for the "museum sulfates" above, the modal abundance of $Fe_7S_8$ cannot account for all the sulfur measured in these meteorites; neither can adding sulfur bound to pentlandite, $(Fe,Ni)_9S_8$. There is another sulfur-bearing phase in CM chondrites that is frequently ignored and should be considered as surface or interior phase together with other hydrated silicates on the moons.

**Tochilinite** $(Fe^{2+})_{5-6}(Mg,Fe^{2+})_5S_6(OH)_{10}$ belongs to the groups of hydroxysulfides and hydrated sulfides. It is an alteration product of magnetite in serpentinites and on Earth and CM-chondrites it is frequently associated with calcite $CaCO_3$, dolomite $CaMg(CO_3)_2$, brucite $Mg(OH)_2$, and magnetite $(Fe_3O_4)$. In some terrestrial occurrences it may contain $Fe^{3+}$. Its habits of hexagonal plates somewhat looks like hexagonal plates of pyrrhotite $Fe_7S_8$. In CM chondrites, it is known in close associations with calcite but for most, it is finely intergrown with serpentine and was known as that the "Fe-S-O phase" until it was identified (see Lodders and Fegley 2011 for more details and references).

**Green-rust:** $[Fe^{2+}_{6-x}Fe^{3+}_x(OH)_{12}]^{x+}\ [(A^{n-})_{x/n} \cdot yH_2O]^{x-}$

Metal and sulfides have particular oxidation paths in anoxic conditions which is relevant to carbonaceous chondrites and to the aqueous alteration products that occurred in the interiors of for the outer planet moons. The intermediate products of metal and sulfide corrosion in controlled anoxic neutral and basic solutions are collectively called "green rust". Some of the material here is drawn from our book (Lodders & Fegley 2011) and we refer the reader to this for more details and references.

The natural mineral, fougerite $(Fe^{2+},Mg)_6Fe^{3+}_2(OH)_{18}\cdot 4H_2O$, resembles green rusts and they are all are unstable in air on Earth because they contain $Fe^{2+}$ with strong reducing action. Green rust contains $Fe^{2+}$ and $Fe^{3+}$ in hydroxide layers that are stabilized by inter-layers made of negatively charged anions. The general green rust formula is

$[Fe^{2+}_{6-x}Fe^{3+}_x(OH)_{12}]^{x+}\ [(A^{n-})_{x/n}\cdot yH_2O]^{x-}$

where x = 0.9 to 4.2. The $(A^{n-})_{x/n}$ stands for a n-valent anion such as $Cl^-$, $SO_4^{2-}$, $CO_3^{2-}$, $OH^-$, $HCO_3$, $SO_3^{2-}$, $Br^-$ or $I^-$. Sulfide $S^{2-}$ has not yet been reported as a stabilizing anion. Green rusts appear as oxidation products with intermediate valence states between $Fe^{2+}$ and $Fe^{3+}$. The end-member for x = 0 is "6 $Fe(OH)_2\cdot yH_2O$", which is the first expected oxidation product from Fe metal in aqueous solution with strong kinship to ferrihydrite.

In solution with carbonates, $FeCO_3$ (siderite) formation from metal competes with the formation of $Fe(OH)_2$ (ferrihydrite). Green rust intermediates facilitate $Fe^{2+}$ to $Fe^{3+}$ oxidation to magnetite, known as the Schikorr reaction (Schikorr 1933), written simplified:

$$Fe_6(OH)_{12}CO_3 + OH^- = 2\ Fe_3O_4\ HCO_3^- + 5H_2O + H_2$$

In magnetite, the $Fe^{2+}$ to $Fe^{3+}$ ratio is 1:2, and at least two $Fe^{3+}$ must be present per formula of green rust to make one $Fe_3O_4$. This corresponds to x = 2 in the green rust formula

$[Fe^{2+}_4Fe^{3+}_2(OH)_{12}]^{2+}[CO_3\cdot yH_2O]^{2-}$ (green rust)

$\quad = Fe_3O_4$ (magnetite)+ 3 $Fe(OH)_2$ (solid) + $CO_2$ (aq) + (3+y)$H_2O$ (liq)

Oxidation of all initial Fe-metal to magnetite requires a nominal x = 4, close to the known limit of x in the green rust formula:

$[Fe^{2+}_2Fe^{3+}_4(OH)_{12}]^{4+}[2CO_3\cdot(2-4)\ H_2O]^{4-}$ (green rust)

$\quad = 2\ Fe_3O_4$ (magnetite) + 2 $CO_2$ (aq) + (6+y) $H_2O$ (liq)

In $CO_2$-bearing solutions, siderite forms

Fe (metal) + 4 $CO_2$ (aq) + $H_2O$ (liq) = $FeCO_3$ (siderite)+ 2 CO (gas)

We propose green rusts as a mechanism to oxidize the metal phase during accretion of the moons when the rocky materials are in contact with the aqueous environment. It effectively oxidizes metallic iron and moves iron into magnetite and frees sulfur from initial troilite. Green rusts could also be of interest as phases at the contact zones to subsurface oceans or ice layers. We also find that there are parallels of green rusts to tochilinite, which need to be investigated further as both types of minerals are highly relevant to anoxic conditions of the moons during their formation and at ocean-silicate interfaces.

## Conclusions/Implications

Given the importance of $NH_4SH$ condensation for condensation cloud models on Jupiter, Saturn, Uranus, and Neptune it is surprising to realize the equilibrium constant (and thus the standard Gibbs energy change) for $NH_4SH$ condensation is based upon a few vapor pressure measurements from 1881 – 1932. There are no experimental measurements of the heat capacity (and thus entropy) of $NH_4SH$, its enthalpy of formation at 298.15 K, or of its dissociation pressure at temperatures above 44.4 °C (≈317.6 K). Values for the thermodynamic properties of $NH_4SH$ at 298.15 K (e.g., Wagman et al. 1982) are based upon the old vapor pressure data. It is important to test the validity of the thermodynamic data for $NH_4SH$ formation from ammonia and hydrogen sulfide by new experimental measurements. We suggest three important questions that should be considered. First, are $NH_3$ and $H_2S$ the only two gases in the saturated vapor above $NH_4SH$ or is $NH_4SH$ also present, especially at higher temperatures and dissociation pressures? Second, how does the vapor pressure curve and saturated vapor composition change as the melting point of $NH_4SH$, (≈393 K, Scheflan and McCrosky 1932), is approached. This latter question is especially important for modeling $NH_4SH$ cloud condensation in the deep atmospheres of Uranus and Neptune.

Separately, but no less important, future entry probes should be designed so cloud particles in the region of the $NH_4SH$ cloud can be analyzed to confirm the cloud is $NH_4SH$ instead of another $NH_3 – H_2S$ bearing phase.

As mentioned earlier, there are no observations of HS, $H_2S_2$, or the sulfur allotropes on any of the giant planets and the $H_2S$ photochemical scheme proposed for these planets remains theoretical. One possibility is to measure the chemical scale height of $H_2S$ in the photochemically active region of one of the giant planets (presumably Jupiter where this would be easiest to do) and compare it to the predicted chemical scale height from the photochemical scheme. A measurement of atmospheric vertical mixing in the same region would also be required. Direct measurement of the reactive photochemical intermediates may be possible on a suitably equipped entry probe or pair of entry probes to allow atmospheric spectroscopy from one probe to another. Perhaps one of the methods used to measure the hydroxyl radical in Earth's atmosphere could be adapted for use on Jupiter.

For the moons, we summarized the sulfur cycles for Io and the icy Galilean moons. Many open questions remain about the exact nature of sulfur-bearing compounds, especially that of sulfur containing salts. Considering various cosmochemical constraints, we suggest that in addition to pyrrhotite, tochilinite and green rusts could be important sulfur bearing compounds among hydrous silicates of the moons. Important questions include the nature of S-bearing green rusts on outer planet satellites, whether tochilinite is a sink for sulfur, and the exact relationships between tochilinite and green rust. Sulfate-bearing green rusts are known but sulfide-bearing green rusts are unknown but could perhaps exist on one or more satellites. The presence of phosphorus (as sodium phosphate) but not sulfur in grains emitted by plumes on Enceladus is also intriguing (Postberg et al. 2023). In nitrogen-carbon rich worlds such as Titan and sulfates are unstable but sulfides such as $NH_4SH$ and possibly thiazyl compounds could be important, which needs more exploration. Unfortunately, nothing is known about the sulfur chemistry on the very cold Uranian and Neptunian moons but their ice composition could reveal more information about the sulfur speciation within the early solar system.

## Acknowledgments

BF thanks J.S. Lewis for introducing him to Jovian atmospheric chemistry and the world of icy satellites when he was an undergraduate and graduate student at MIT in the 1970s. We thank Brother Guy Consolmagno and Jeff Kargel for their helpful reviews. Work supported by NSF-AST 2108172 and by the McDonnell Center for the Space Sciences.